\documentclass[prb, superscriptaddress,aps, twocolumn, showpacs, bibliography, footnoteinbib]{revtex4-2}
\pdfoutput=1 
\usepackage{graphicx}
\usepackage[justification=raggedright]{caption}
\usepackage{amsmath, amsfonts, amssymb, bm}
\usepackage{physics}
\usepackage{dsfont}
\usepackage{natbib}
\usepackage{dcolumn}   
\usepackage[mathscr]{eucal}
\usepackage[dvipsnames]{xcolor}
\usepackage{hyperref}
\hypersetup{colorlinks, linkcolor=OrangeRed,citecolor=RoyalBlue,urlcolor=NavyBlue, breaklinks}
\usepackage[all]{hypcap} 
\usepackage{mathtools}
\usepackage{dsfont}
\usepackage{multirow}
\usepackage{hhline}
\usepackage{array}
\newcolumntype{P}[1]{>{\centering\arraybackslash}p{#1}}

\newcommand{\dr}[1]{\delta\rho_{#1}}

\usepackage{soul}
\setstcolor{red}

\begin{document}
\title{Hierarchical hydrodynamics in long-range multipole-conserving systems}
\author{Jacopo Gliozzi}
\affiliation{Department of Physics and Institute for Condensed Matter Theory,\\
University of Illinois at Urbana-Champaign, Urbana, Illinois 61801, USA}
\author{Julian May-Mann}
\affiliation{Department of Physics and Institute for Condensed Matter Theory,\\
University of Illinois at Urbana-Champaign, Urbana, Illinois 61801, USA}
\author{Taylor L. Hughes}
\affiliation{Department of Physics and Institute for Condensed Matter Theory,\\
University of Illinois at Urbana-Champaign, Urbana, Illinois 61801, USA}
\author{Giuseppe De Tomasi}
\affiliation{Department of Physics and Institute for Condensed Matter Theory,\\
University of Illinois at Urbana-Champaign, Urbana, Illinois 61801, USA}
\begin{abstract}
This work investigates the out-of-equilibrium dynamics of dipole and higher-moment conserving systems with long-range interactions, drawing inspiration from trapped ion experiments in strongly tilted potentials. 
We introduce a hierarchical sequence of multipole-conserving models characterized by power-law decaying couplings.  
Although the moments are always globally conserved, adjusting the power-law exponents of the couplings induces various regimes in which only a subset of multipole moments are effectively locally conserved.
We examine the late-time hydrodynamics analytically and numerically using an effective classical framework, uncovering a rich dynamical phase diagram that includes subdiffusion, conventional diffusion, and Lévy flights. Our results are unified in an analytic reciprocal rule that captures the nested hierarchy of hydrodynamics in multipole conserving systems where only a subset of the moments are locally conserved.
Moreover, we extend our findings to higher dimensions and explore the emergence of long-time scales, reminiscent of pre-thermal regimes, in systems with low charge density. Lastly, we corroborate our results through state-of-the-art numerical simulations of a fully quantum long-range dipole-conserving system and discuss their relevance to trapped-ion experimental setups.
\end{abstract}

\maketitle

\section{Introduction}
Recent advances in controlled experimental platforms, such as ultracold atoms in optical lattices~\cite{Bloch_08,Bloch2012}, trapped ions~\cite{Georgescu_14,Blatt2012}, and superconducting qubits~\cite{Krantz_19,Kjaergaard_20,Siddiqi2021}, have sparked considerable interest in the out-of-equilibrium dynamics of isolated quantum many-body systems. Such systems can display a wide range of fascinating and unexpected behaviors, including the emergence of new phases of matter ranging from exotic topological phases to time crystals~\cite{Polko_2011,Lindner2011,Wilczek_12,Khemani_16,Else_17,Yao_17,Oka_2019}.

Generic quantum systems are expected to thermalize, meaning that their long-time steady states are described by a finite number of global conserved quantities, such as energy, particle number, or charge. When such systems evolve, any non-equilibrium dynamics ultimately lead to an equilibrium thermal state~\cite{Deutsch_91, Srednicki_94, Rigol2008, Rigol_12, Nandkishore_15,DAlessio_16,Abanin_18}. Indeed, any initial inhomogeneities of the globally conserved quantities are smoothed out at late times by non-equilibrium transport. These processes can be described in an effective classical hydrodynamic framework that emerges in interacting quantum systems~\cite{Mukerjee_06,Lux_14,Medenjak_17,Rakovszky2018,Khemani2018,Wurtz_20,Zonaci_21}. For example, the long-time dynamics of a system with U(1) charge conservation exhibit diffusive relaxation.

Here we are interested in systems that conserve both a global U(1) charge and one or more of its higher multipole moments, e.g., dipole or quadrupole moments. Fundamentally new equilibrium~\cite{griffin2015scalar, pretko2018fracton, seiberg2020field, he2020lieb, dubinkin2021lieb, dubinkin2021theory, you2021multipolar, may2021topological,Lake2022a} and out-of-equilibrium phenomena~\cite{prem2017glassy, Sala2020,Rakovszky2020,Khemani2020,Gromov2020,Feldmeier2020, Morningstar2020, Feldmeier2021, Zechmann2022, Pozderac23} emerge in such systems.  For instance, dipole-conserving systems can exhibit anomalously slow dynamics due to changes in Fick's law, which lead to modified diffusion equations for the charges~\cite{Feldmeier2020, Gromov2020}. Indeed, dipole conservation  leads to dynamical constraints reminiscent of fractonic systems ~\cite{Nandkishore_19, vijay2015new, vijay2016fracton, pretko2017subdimensional, pretko2018fractonE,  pretko2020fracton}. For Hamiltonians having sufficiently short-range interactions, these dynamical constraints can cause strong Hilbert space fragmentation, where the Hilbert space splits into exponentially many disconnected sectors, and where the number of states in the largest sector is still a vanishing fraction of the full Hilbert space dimension~\cite{GDT_19,Sala2020,Moudgalya_2022,Moudgalya_Motrunich_2022}. As a result, any initial charge configuration is constrained to explore only a small portion of the Hilbert space, preventing the system from thermalizing.

In comparison, dipole-conserving systems having extended, but still local, interactions  exhibit a weak form of Hilbert space fragmentation~\cite{Pai_2019, Sala2020, Khemani2020, Pozderac23}, meaning that they remain ergodic despite the existence of a measure zero subset of non-ergodic states, dubbed quantum scars~\cite{Kormos2017,Vafek_2017,Moudgalya_18,Moudgalya_scars_18,Turner_18, Moudgalya_2022, Serbyn2021}. For special, fine-tuned initial states, these scar states can dominate the dynamics, leading to highly non-ergodic time evolution with persistent oscillations. However, for generic initial states, these systems thermalize under unitary time evolution, and conserved quantities determine their universal long-time behavior. Such local Hamiltonians can also be tuned from weakly to strongly fragmented by changing the charge density, driving the so-called freezing transition~\cite{Morningstar2020, Feldmeier2021, Pozderac23}.

\begin{figure*}[t]
\centering
\includegraphics[width=0.85\linewidth]{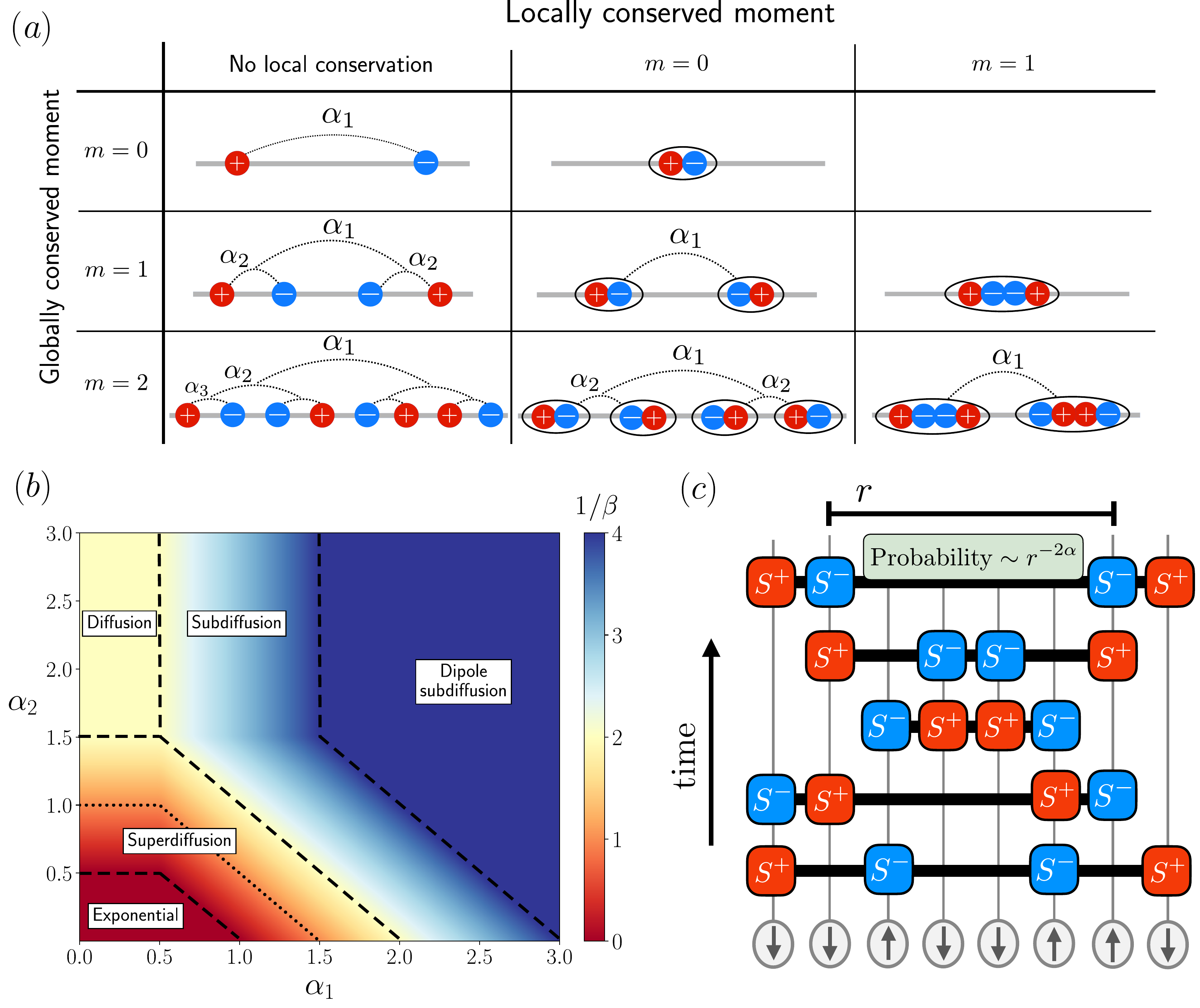}
\caption{(a) 
Table summarizing the hierarchy of long-range multipole conserving models. A system can conserve all moments up to  the $m$th moment globally, but for finite long-range exponents $\alpha_i$, this conservation may not be valid at the local level. Taking the relevant short-range limit $\alpha_i \rightarrow \infty$, we recover local conservation. (b) Phase diagram for the long-time dynamics of a dipole-conserving system with general long-range interactions. Here $\alpha_1$ and $\alpha_2$ are the two exponents controlling the range of interaction between and within dipoles, respectively. For large $\alpha_1$ we recover dipole subdiffusion typical of the short-range case, and longer-range systems can also exhibit regimes of diffusion, superdiffusion (including the special case of ballistic transport, marked with a dotted line), and exponential relaxation which indicates a breakdown of hydrodynamics (c) An example of a dipole-conserving cellular automaton circuit. Gates are applied with a probability that decays algebraically with their range, and product states are mapped to product states.}
\label{fig:1}
\end{figure*}

In realistic systems, dipole conservation can be effectively imposed at low energies by a large linear potential~\cite{Refael2019,Taylor_20,Guardado_2020}. The long-time dynamics of dipole-conserving systems can therefore be explored in quantum platforms with a strong external ``tilt," and recent experimental realizations include ultracold atoms~\cite{Scherg2021,Kohlert223,Guardado_2020,DeMarco2022}, superconducting qubits~\cite{Gou_21}, and trapped-ion systems~\cite{Morong2021}. For example, Refs.~\onlinecite{Scherg2021,Kohlert223} probed the strong fragmentation regime of a tilted one-dimensional Fermi-Hubbard model, while Ref.~\onlinecite{Guardado_2020} investigated the anomalous relaxation of charge in the two-dimensional tilted Fermi-Hubbard model. In many such platforms, the interactions are long-ranged and decay algebraically with distance, e.g., the Coulomb potential between trapped ions or the dipole-dipole interactions in Rydberg atom arrays~\cite{Defenu2021}. However, previous theoretical works on out-of-equilibrium dynamics in the presence of dipole conservation have primarily considered short-range interactions.

In this work, we investigate  dipole- and higher-moment-conserving systems subjected to long-range, algebraically decaying interactions. This is partially motivated by the aforementioned experimental study of trapped ions in a strong tilted field~\cite{Morong2021}. As algebraically decaying interactions are inherent in such systems, it is natural to ask how the transport of charges is affected by long-range interactions when multipole moments are conserved. Quantum simulators based on trapped ions~\cite{Islam2011, Morong2021, Jurcevic_17, Smith2016,Hess_17, Morong2021} also provide an ideal setting to realize the predictions of our work. 

Our work builds upon previous theoretical~\cite{Schuckert2020} and experimental~\cite{Joshi2022} works, which considered hydrodynamic descriptions of trapped-ion setups with U(1) charge conservation and long-range interactions. In particular, Ref.~\onlinecite{Schuckert2020} established the existence of three dynamical regimes depending on the range of the interactions: (i) a universal diffusive transport regime, (ii) an intermediate regime with emergent non-local hydrodynamics, and (iii) a ``super"-long-range regime in which hydrodynamics breaks down. Building upon these results, here we establish a sequence of hierarchical models that conserve higher-moments, such as dipole and quadrupole moments, and we choose the interactions to be algebraically decaying couplings controlled by a set of exponents $\{\alpha_i\}$, as illustrated in Fig.~\ref{fig:1}~(a). For instance, the second row of Fig.~\ref{fig:1}~(a) displays couplings that conserve dipole moment, which we characterize by two exponents, $\alpha_1$ and $\alpha_2$. The former governs the typical interaction between dipoles, $(+,-)\longleftrightarrow (-,+)$, while the latter modulates the interaction between the constituent charges of each dipole, $(+) \longleftrightarrow (-) $. 
For small values of $\alpha_1$ and $\alpha_2$, the model becomes highly non-local, and neither charge nor dipole moment is \emph{locally} conserved. However, as $\alpha_2$ tends to infinity, charge becomes locally conserved, and compact dipoles $(+,-)$ can propagate over long distances. Finally, when both $\alpha_1$ and $\alpha_2$ tend towards infinity we recover the short-range case, where both charge and dipole moment are locally conserved. 

By iterating this construction, we obtain a sequence of models that globally conserve all moments up to the $m$th multipole moment.  Employing both analytical and numerical techniques, we study the late-time hydrodynamics of these models using an effective classical description. Our results reveal a rich phenomenology, and the out-of-equilibrium phase diagrams of each model host various dynamical regimes, including conventional diffusion, Lévy flights, and stable subdiffusion. As an example, in Fig.~\ref{fig:1}~(b) we show the phase diagram for the dipole case with exponents $\alpha_1$ and $\alpha_2$ (we will discuss this in more detail below). In this phase diagram we find that at large $\alpha_1$ and $\alpha_2$, the system behaves like short-ranged dipole-conserving systems and exhibits subdiffusion. 
As we lower $\alpha_1$, which controls the typical interaction scale between dipoles, we obtain conventional diffusion, like short-ranged charge conserving systems. 
More generally, we find that in models that globally conserve up to $m$th multipole moments, we can understand the various dynamical regimes as the dynamics of systems where only the subset of $s$th moments, with $s\le m$, are both globally and \emph{locally} conserved. 
For example, quadrupole-conserving models feature hydrodynamic regimes that describe locally conserved quadrupoles (and all lower moments), other regimes where the dynamics is governed by locally conserved dipoles, and, finally, regimes where only charge is effectively locally conserved.

We also investigate the hydrodynamics of a system that has long-range, dipole-conserving interactions that match the effective interactions of trapped-ion systems subject to strong tilted fields~\cite{Guardado_2020}. Furthermore, we extend our work to higher dimensions and discuss prethermal-like regimes where the dynamics is nearly frozen for systems at low particle density. 
Finally, we support our results by performing state-of-the-art numerical modeling of the time evolution of a fully quantum, long-range dipole-conserving model.

The remainder of our work is organized as follows: In Section~\ref{Sec:model}, we define our multipole models and their hierarchical structure. In Section~\ref{sec:Methods}, we describe the methods used in our numerical simulations, such as cellular automata circuits and dynamical probes. Section~\ref{sec:Results} serves as the core of our work, where we present our analytical derivations for the hydrodynamics of long-ranged systems with conserved multipole moments and support them with clear numerical evidence. Finally, in Section~\ref{sec:Discussion} we discuss the experimental significance of our results and offer some concluding remarks.

\section{Models}~\label{Sec:model}
\subsection{Dipole-conserving model}\label{ssec:DipoleConservingModel}
To study the dynamics of dipole-conserving systems with long-range couplings, we consider the following model of local spin-$S$ degrees of freedom in one spatial dimension,
\begin{equation}
    H = \sum^{'}_{i,j,n}\left( J_{i,j,n} S^+_i S^-_{i+n} S^-_j S^+_{j+n}  + \text{h.c.}\right),
\label{eq:GeneralDipoleConservingHam}
\end{equation}
where $S^{\pm}_i  = (S^x_i \pm i S^y_i)/2$ are the raising and lowering operators for the spin at site $i$. The primed sum indicates that the sum is taken over values of $i,j,n$ such that the site indices on the spin are ascending, ($i<i+n<j<j+n\le L$), where $L$ is the length of the system, and we use open boundary conditions. This Hamiltonian conserves both the charge $P^{(0)}=\sum_i S^z_i$ and the dipole moment $P^{(1)} = \sum_i i S^z_i$. Since we are dealing with spin systems where the $U(1)$ charge is the $z$-component of the spin, we will use ``charge" and ``magnetization" synonymously. The $J_{i,j,n}$ term can be viewed as a dipole hopping term, where the first two operators, $S^+_i S^-_{i+n} $, create a dipole of length $n$ anchored at site $i$, and the second two operators, $S^-_j S^+_{j+n}$, remove a dipole of length $n$ anchored at site $j$. 

In this work, we will consider two forms of $J_{i,j,n}$. First, 
\begin{equation}
    J_{i,j,n} =  \frac{J_0}{|i-j|^{\alpha_1} |n|^{\alpha_2}},
\label{eq:alpha1alpha2}\end{equation}
where $J_0$ is a constant, and $\alpha_1$ and $\alpha_2$ are free parameters. We will refer to Eq.~\eqref{eq:GeneralDipoleConservingHam} with this form of $J_{i,j,n}$ as the $\alpha_1,\alpha_2$-model (see Fig.~\ref{fig:1}~(a)).
Interpreting $J_{i,j,n}$ as a dipole hopping term, $\alpha_2$ controls the length of the dipoles, and $\alpha_1$ controls the range of the hopping. As with any model with algebraically long-ranged interactions, the energy becomes superextensive below some threshold values of the long-range exponents $\{\alpha_i\}$ where certain integrals diverge as $L\rightarrow \infty$~\cite{Kac1963}. To avoid pathological behavior in the thermodynamic limit, we must correct the Hamiltonian by dividing by a so-called Kac factor $\mathcal{N}(L, \alpha_i)$, which makes the energy extensive.
In the case of the coupling in Eq.~\eqref{eq:alpha1alpha2}, the relevant superextensive regimes are $\alpha_1 < 1/2$ and $\alpha_1+\alpha_2 < 1$, and we henceforth assume that such a correcting factor is implicitly included when needed.

It will be useful to discuss some important limits of the $\alpha_1,\alpha_2$-model. In the limit where $\alpha_2 \rightarrow \infty$, the $\alpha_1,\alpha_2$-model becomes 
\begin{equation}
     H = \sum^{'}_{i,j}\left( \frac{J_0 }{|i-j|^{\alpha_1}} S^+_i S^-_{i+1} S^-_j S^+_{j+1} + \text{h.c.}\right),
\label{eq:LongRangeModel1}\end{equation}
where the primed sum again indicates that sum is taken over $i$ and $j$ such that the site indices on the spins are ascending ($i<i+1<j<j+1$). This Hamiltonian can be considered as a model of long-range hopping for short dipoles with a length of one lattice spacing. In the limit where $\alpha_1 \rightarrow \infty$, the $\alpha_1,\alpha_2$-model becomes local
\begin{equation}
     H = \sum_{i}\left( J_0  S^+_i S^-_{i+1} S^-_{i+2} S^+_{i+3} + \text{h.c.}\right),
\label{eq:ShortRangeModel}\end{equation}
regardless of the value of $\alpha_2$ \footnote{In our parametrization of the long-range exponents, $\alpha_1$ is dominant over $\alpha_2$ and controls the overall range of the model. A more symmetric choice, like $J_{i,j,n} \sim |i-j+n|^{-\alpha_1} |n|^{-\alpha_2}$, yields independent exponents, but obscures the hierarchical nature of multipole-conserving hydrodynamics.}. In the opposite limit, where $\alpha_1 \rightarrow 0$, the Hamiltonian is 
\begin{equation}
     H = \sum^{'}_{i,j,n}\left( \frac{J_0}{|n|^{\alpha_2}} S^+_i S^-_{i+n} S^-_{j} S^+_{j+n}  + \text{h.c.}\right).
\label{eq:LongRangeModel2}\end{equation}
This Hamiltonian consists of an all-to-all hopping of dipoles, where the length of the dipoles is still controlled by $\alpha_2$. 
In the limit where both $\alpha_1 \rightarrow 0$ and $\alpha_2 \rightarrow 0$ the model has an all-to-all form
\begin{equation}
     H = \sum^{'}_{i,j,k,l}\left( J_0  S^+_i S^-_{j} S^-_{k} S^+_{l} \ \delta_{i-j,k-l} + \text{h.c.}\right),
\label{eq:SYK}\end{equation}
which is reminiscent of a four-body, SYK-like interaction~\cite{Sachdev_93, Kitaev2015} with an additional center-of-mass constraint.

The second form of $J_{i,j,n}$ in Eq.~\eqref{eq:GeneralDipoleConservingHam} that we consider is the following:
\begin{equation}\begin{split}
   J_{i,j,n} = & \frac{J_0}{|i-j|^{1+\gamma} |n|^{1+\gamma}} \left [\frac{1}{|i-j+n|^\gamma}-\frac{1}{|i-j-n|^\gamma} \right ],
\end{split}\label{eq:JTilt}\end{equation}
 where $\gamma$ is a free parameter. We will refer to Eq.~\eqref{eq:GeneralDipoleConservingHam} with this form of $ J_{i,j,n}$ as the tilted Hamiltonian since, as we show in Appendix \ref{app:EffHamiltonian}, this term appears in the effective Hamiltonian for a long-range XY spin chain in a tilted potential, analogous to trapped-ion platforms~\cite{Morong2021}.

Some short-ranged versions of the models we consider have been shown to exhibit Hilbert space fragmentation and ergodicity breaking~\cite{Sala2020, Sala2021, Khemani2020}. The addition of long-range interactions ensures that even weak fragmentation, where exponentially many non-thermal eigenstates remain, is avoided. We nevertheless primarily restrict our attention the Hilbert space sector with vanishing magnetization and dipole moment, which is the largest, and therefore the most typical, sector.

\subsection{Quadrupole-conserving model
}The dipole-conserving model in Eq. \eqref{eq:GeneralDipoleConservingHam} can be generalized to a quadrupole-conserving model,
\begin{equation}\begin{split}
    H = \sum^{'}_{i,j,n_1,n_2} Q_{i,j,n_1,n_2} &\Big[S^+_i S^-_{i+n_1} S^-_{i+n_2} S^+_{i+n_1+n_2} \\ &\times S^-_{j} S^+_{j+n_1} S^{+}_{j+n_2} S^-_{j+n_1+n_2} + \text{h.c.}\Big],
\end{split}\label{eq:GeneralQuadrupoleConservingHam}\end{equation}
where the primed sum again indicates that the sum is taken such that the site indices on the spins are ascending ($i<i+n_1<i+n_2<i+n_1+n_2<j...$). 
This system preserves the charge, the dipole moment, and the quadrupole moment $P^{(2)}=\sum_{j} j^2 S^z_j$. This is not the most general form of a quadrupole-conserving Hamiltonian, but it is sufficient to discuss the connections between dipole and quadrupole-conserving physics. The $Q_{i,j,n_1,n_2}$ term can be considered to be a quadrupole hopping term where the first four terms, $S^+_i S^-_{i+n_1} S^-_{i+n_2} S^+_{i+n_1+n_2}$, create a quadrupole anchored at site $i$ and the second four terms, $S^-_{j} S^+_{j+n_1} S^{+}_{j+n_2} S^-_{j+n_1+n_2}$, remove a quadrupole anchored at site $j$. These quadrupoles are composed of two oppositely oriented dipoles of length $n_1$ that are separated by a distance $n_2$. 

Analogously to the $\alpha_1, \alpha_2$-model, we will consider the following form of $Q_{i,j,n_1,n_2}$:
\begin{equation}
\label{eq:Q_coefficient}
    Q_{i,j,n_1,n_2} = \frac{Q_0}{ |i-j|^{\alpha_1} |n_1|^{\alpha_2}  |n_2|^{\alpha_3} }.
\end{equation}
We will refer to Eq.~\eqref{eq:GeneralQuadrupoleConservingHam} with this form of $Q$ as the $\alpha_1,\alpha_2,\alpha_3$-model. If we interpret $Q_{i,j,n_1,n_2}$ as a quadrupole hopping term, $\alpha_1$ controls the distance over which the quadrupoles hop. The value of $\alpha_2$ controls the distance between the dipoles that make up the quadrupoles, and  $\alpha_3$ controls the length of the constituent dipoles. This should be compared to the dipole-conserving $\alpha_1,\alpha_2$-model, where $\alpha_1$ controls the distance over which the dipoles hop, and $\alpha_2$ controls the distance between the charges that make up the dipole. This structure can be directly extended to an $m$th moment-conserving model, where $\alpha_1$ controls the distance over which $m$-poles hop. The $m$-poles are composed of two $(m-1)$-poles separated by a distance controlled by $\alpha_2$, The $(m-1)$-poles are composed of two $(m-2)$-poles separated by a distances controlled by $\alpha_3$, and so forth. 

Let us now mention some relevant limits of the $\alpha_1,\alpha_2,\alpha_3$-model. When $\alpha_1 \rightarrow \infty$,  Eq. \eqref{eq:GeneralQuadrupoleConservingHam} becomes a local quadrupole-conserving system
\begin{equation}\begin{split}
    H = &\sum_{i}Q_0  \left [S^+_i S^-_{i+1} S^-_{i+2} S^+_{i+3} S^-_{i+4} S^+_{i+5} S^{+}_{i+6} S^-_{i+7}  + \text{h.c.} \right ]. 
\end{split}\label{eq:GeneralQuadrupoleLocal}\end{equation}
In the limit where ${\alpha_1 \rightarrow 0}$, Eq. \eqref{eq:GeneralQuadrupoleConservingHam} is 
\begin{equation}\begin{split}
    H = \sum^{'}_{i,j,n_1,n_2} \frac{Q_0}{{|n_1|^{\alpha_2} |n_2|^{\alpha_3}}} &\Big[S^+_i S^-_{i+n_1} S^-_{i+n_2} S^+_{i+n_1+n_2} \\ &\times S^-_{j} S^+_{j+n_1} S^{+}_{j+n_2} S^-_{j+n_1+n_2} + \text{h.c.} \Big],
\end{split}\label{eq:GeneralQuadrupoleAllToAll}\end{equation}
which consists of an all-to-all hoping of quadrupoles. The quadrupoles are composed of two dipoles (with dipole lengths controlled by $\alpha_3$) that are separated by a distance controlled by $\alpha_2$. The interpretation of other limits involving $\alpha_2$ and $\alpha_3$ follows from our previous discussion in Sec. \ref{ssec:DipoleConservingModel}. 

\section{Methods}\label{sec:Methods} 
To understand the late-time out-of-equilibrium dynamics governed by the conserved quantities of the underlying system, we efficiently simulate time evolution using a classical cellular automaton that respects the same conservation laws~\cite{Sollich_03,Redner2010,Medenjak_17,Gopalakrishnan_2018, Iaconis2019,Feldmeier2020,Schuckert2020} (see Fig.~\ref{fig:1}~(c)). This approach simplifies the dynamics such that all unitary time-evolution operators commute. As a result, cellular automata map product states to product states and effectively transform the time evolution of a spin system into a series of spin-flips. Although such spin-flip gates do not generate state entanglement, they can still capture operator spreading in chaotic quantum many-body systems and their hydrodynamics~\cite{Iaconis2019}. The main assumption for the applicability of this approach is that the system is ergodic and hence, that it thermalizes. In the ergodic regime, the long-time dynamics are believed to be dominated by emergent classical hydrodynamics. More precisely, the scaling behavior of correlation functions is universal and characterized by the hydrodynamics, while quantum fluctuations enter as non-universal coefficients. We consider an infinite temperature ensemble of initial quantum states to capture the relaxation of a localized excitation in a thermal bath. As a result, quantum fluctuations are effectively washed out, and classical cellular automaton simulations yield the same hydrodynamics as full quantum time evolution~\cite{Schuckert2020, Feldmeier2020, Iaconis2019, Burchards2022}.

Hence, to evolve the system we perform a series of gates that update the spin configuration at each time-step. The update gates are chosen to respect the symmetry of the original Hamiltonian in Eq.~\eqref{eq:GeneralDipoleConservingHam}, and the transition rate between product states is dictated by Fermi's golden rule:
\begin{equation}\label{eq:fermi}
W_{s',s} = \abs{\mel{s_1',s_2', \ldots}{H}{s_1,s_2, \ldots}}^2,  
\end{equation}
where $|s\rangle = |s_1,s_2,\cdots \rangle$ is a product state in the $S^z$-basis, and $s_i\in \{-S, -S+1,\cdots, S-1, S\}$. For example, in the case of dipole-conserving models like Eq.~\eqref{eq:GeneralDipoleConservingHam}, an update gate involving four site indices $\{i,j,k,l\}$, such as $U_{i,j,k,l}\sim S^+_i S^-_{j} S^-_k S^+_{l}$ ($U_{i,j,k,l}\sim S^-_i S^+_{j} S^+_k S^-_{l}$), is applied with a probability proportional to the distance between the sites, i.e., $W\propto  |J_{i,j,k,l}|^2 \delta_{l-i,k-j}$. In Fig.~\ref{fig:1}(c) we show a schematic example of such a dipole-conserving circuit architecture.

A standard description of out-of-equilibrium dynamics employs the infinite-temperature, connected spin-spin correlation function
\begin{equation}
\label{eq:correlator}
C(|i-j|, t) = \langle S_i^z(t) S_j^z(0) \rangle -\langle S_i^z(t) \rangle \langle S_j^z(0) \rangle,
\end{equation}
where $\langle \cdots \rangle$ indicates an average over random initial spin configurations in the $S^z$ basis. We thus expect that we are probing the largest and most typical Hilbert space sectors, i.e., those in which $P^{(m)}\approx 0$. From this correlation function we can consider the dynamical exponent $\beta,$ which quantifies spin relaxation, and determines the long-time asymptotic behavior of the return probability $C(x=0, t) \sim t^{-\beta}$. 
For short-ranged systems with only U(1) charge conservation (i.e. systems with conserved total charge/magnetization $\sum_j S_j^z$), the spread of charge is diffusive. 
In the diffusive regime, the long-wavelength limit of the correlator is Gaussian: $C(x,t) \sim e^{-x^2/{4Dt}}/\sqrt{D t}$, or in momentum space, $C(k,t)\sim e^{-4 Dk^2 t}$, with $D$ the diffusion constant. At equal distances, $C(x=0,t)\sim t^{-1/2}$, and therefore $\beta = 1/2$~\cite{Mukerjee_06,Bera_17,Schuckert2020}.

In a charge-conserving system, diffusion remains stable in the presence of long-range, algebraically decaying couplings $J_{i,j} \sim 1/{|i-j|^\alpha}$ for $\alpha >3/2$, as shown in Ref.~\onlinecite{Schuckert2020}. However, for longer-range interactions the system exhibits superdiffusion and eventually a breakdown of the hydrodynamic picture:
\begin{equation}\label{eq:beta_m0}
\beta^{(m=0)}(\alpha) = 
\begin{cases}
1/2 & \alpha > 3/2 \\
{1}/({2\alpha-1}) & 1/2 \leq \alpha \leq 3/2 \\
\infty & \alpha < 1/2,
\end{cases}
\end{equation}
where we have used the notation $\beta^{(m=0)}$ to indicate the dynamical relaxation exponent in systems that only conserve the zeroth moment of a U(1) charge.
For $\alpha<1/2$, hydrodynamics is no longer applicable and the charge relaxes exponentially with a characteristic timescale set by the system size through the (implicit) Kac factor. 

In general, diffusion is associated with the existence of a mean free path $\ell_{mf}$, which leads to an effective random-walk description of the dynamics. Thus, the correlator in Eq.~\eqref{eq:correlator} takes the form of a Gaussian. The stability of the diffusive phase in U(1) charge conserving systems for $\alpha>3/2$ can be derived by applying the central limit theorem~\cite{Redner2010, Schuckert2020}. Indeed, in this regime the second moment of the hopping-rate probability distribution $W_{i,j}\sim 1/{|i-j|^{2\alpha}}$ is finite, and therefore its mean free path is as well. 

In contrast, the dynamics for $1/2<\alpha<3/2$ are described by Lévy flights~\cite{Zaburdaev_15}, which are dominated by large fluctuations. As a result, particles typically scatter in short jumps, but once in a while, with a rare but finite probability, they can undergo a macroscopic jump of order $L^{(3-2\alpha)/2}$. Phenomenologically, this may be interpreted as giving rise to a time-dependent mean-free path. The resulting $\ell_{mf}$ can be found using the ``extremal criterion'' from extreme value statistics~\cite{Redner2010}, 
\begin{equation}
\label{eq:Levy}
    \int_{\ell_{mf}}^\infty \frac{dr}{r^{2\alpha}} \sim \frac{1}{N \delta t},
\end{equation}
where $N$ is the number of steps taken, and $\delta t$ is the unit of time. 
This criterion gives the expected length of the largest of $N$ steps, which dominates the dynamics and therefore gives an estimate of the mean free path: $\ell_{mf}\sim (N\delta t)^{1/(2\alpha-1)}$.
Moreover, the second moment after $N$ steps is given by $N\int^{\ell_{mf}} dr r^2/r^{2\alpha}\sim N N^{\frac{3-2\alpha}{2\alpha-1}}=N^{2/(2\alpha-1)},$ and therefore $C(0, N \delta t) \sim (N \delta t)^{-1/(2\alpha-1)}$. For the special case of $\alpha =1$, the charge undergoes ballistic transport, and the correlator takes the form of a Lorentzian function $C(x,t)\sim t/(t^2+(\lambda x/t))^2$. From this analysis, we see that in the regime $\alpha<1/2$, the probability rate $W_{i,j}\sim 1/{|i-j|^{2\alpha}}$ is not even normalizable, indicating the breakdown of hydrodynamics.

Short-range models with higher-moment conservation, i.e., $P^{(m)} = \sum_j j^{m} S_j^z$ for $m\ge 1$, exhibit out-of-equilibrium dynamics characterized by anomalously slow transport, e.g.,  $C(k,t) \sim e^{-D k^{2m+2} t}$. This, in turn, implies subdiffusion of the underlying U(1) charge, with dynamical exponent $\beta^{(m)} = 1/(2m+2)$~\cite{Feldmeier2020, Gromov2020}. One of our aims is to generalize the above result to the case of long-range interactions, and to find the analogs of Eq. \eqref{eq:beta_m0} for higher moments. 

\section{Results}\label{sec:Results}

\subsection{Long-range hydrodynamics with dipole symmetry} \label{sec:Analytic_dip}

At long times, the quantum dynamics of ergodic dipole-conserving systems becomes mostly insensitive to particularities of the initial state. Instead, the transport of spin is governed by the symmetries of the underlying Hamiltonian, giving way to an effectively classical hydrodynamic description. We can therefore model the evolution of spin using a classical master equation for the local spin density~\cite{Schuckert2020}. Taking spin-1/2 systems for simplicity, we define the spin density at site $i$ as a non-normalized probability density $\rho_i = \expval{S^z_i} + 1/2 \in [0,1]$. 

The master equation then governs the evolution of this local density in accordance with the dipole-conserving constraint, which exclusively allows dipole-exchange processes. The only place the quantum Hamiltonian \eqref{eq:GeneralDipoleConservingHam} enters into this equation is in the rates of dipole exchange through Fermi's golden rule in Eq.~\eqref{eq:fermi}. The full master equation for long-range dipole-exchange processes is given by:
\begin{equation}
    \label{eq:master_dip}
\begin{split}
    \frac{d \rho_i(t)}{dt} = \sum_{j \neq i} \sum_{n=1}^{|i-j|} & \big  [  W^+_{i,j,n} (1-\rho_i)\rho_{i+n}\rho_j(1-\rho_{j+n}) \\
       &-  W^-_{i,j,n}   \rho_i (1-\rho_{i+n})(1-\rho_j)\rho_{j+n}  \\ 
       &-  W^-_{i,j,n}  (1-\rho_{i-n})\rho_{i}\rho_{j-n}(1-\rho_{j}) \\
       &+  W^+_{i,j,n}  \rho_{i-n} (1-\rho_{i})(1-\rho_{j-n})\rho_{j} \big],
\end{split}
\end{equation}
where the first sum runs over a chain of length $L$ with lattice constant $a=1$.
The four terms correspond to the four sets of dipole-conserving processes that either increase or decrease the spin at site $i$, and $W^\pm_{i,j,n}$ are the rates of these processes~\footnote{More precisely, dipole-conserving dynamics are governed by an infinite tower of coupled master equations where the evolution of the one-point function $\rho_i = \expval{S^z_i}$ depends on the two, three, and four-point functions, which in turn depend on higher correlators. Because we work at infinite temperature, however, we can take the mean field limit $\expval{S^z_i S^z_j \ldots} \approx \rho_i \rho_j \ldots$ to decouple them.}. These terms can be viewed as the probability of applying long-range gates like $S^+_i S^-_{i+n} S^-_{j} S^+_{j+n}$ in a cellular automaton simulation of time evolution like in Fig.~\ref{fig:1}~(c). 
At infinite temperature, the rates of opposite processes must be equal to satisfy detailed balance, and Fermi's Golden rule gives
\begin{equation}
\label{eq:rate_golden}
    W^\pm_{i,j,n} \propto |J_{i,j,n}|^2,
\end{equation}
where $J_{i,j,n}$ is the long-range coupling in the underlying dipole-conserving Hamiltonian \eqref{eq:GeneralDipoleConservingHam}. 

Even if the initial conditions of the spin chain are extremely inhomogeneous, the ergodicity of the Hamiltonian ensures that the spin is spread more and more uniformly as $t\rightarrow \infty$, allowing us to expand the spin density into a static, constant background and a small fluctuation: $\rho_i(t) = \overline{\rho} + \dr{i}(t)$. The fluctuation is independent of the background as long as we work in a sector with an extensive number of spin excitations, i.e., near the sector with zero magnetization ($\overline{\rho}=1/2$). 
To study the long-time dynamics, we then linearize the master equation with respect to this fluctuation, effectively describing the evolution of a localized ``lump'' of spin in a background of constant magnetization. The linearized master equation we find is:
\begin{equation}
    \label{eq:master_dip_lin}
    \begin{split}
\partial_t \rho_{i}(t) = -\overline{\rho} (1-\overline{\rho}) \sum_{i \neq j} \sum_{n=1}^{|i-j|} & |J_{i,j,n}|^2  \bigl[ (\rho_{j+n} - 2 \rho_{j} + \rho_{j-n}) \\
   &- (\rho_{i+n} - 2 \rho_i  + \rho_{i-n}) \bigr],
   \end{split}
\end{equation}
where we have used the fact that $\overline{\rho}$ is constant to replace the density fluctuations $\dr{i}$ with the full density. The prefactor $\overline{\rho}(1-\overline{\rho}) = (1/4 - \expval{S^z}^2)$ is always positive and depends on the background magnetization. In particular, it vanishes when the background spin configuration is all up or all down, and is largest for vanishing magnetization. This reflects the fact that there are no dipole-conserving processes allowed starting from the fully magnetized configurations, while the zero-magnetization sector allows many.

Now let us consider the master equation for short-ranged and long-ranged couplings. First, for short-ranged Hamiltonians like Eq.~\eqref{eq:ShortRangeModel}, where $J_{i,j,n} \sim \delta_{i\pm1,j}$, only the nearest neighbor terms in Eq.~\eqref{eq:master_dip_lin} survive, yielding
\begin{equation}
\label{eq:sr_discretized}
      \partial_t \rho_i \propto -   {(\rho_{i+2} - 4\rho_{i+1}+6\rho_i-4\rho_{i-1} +  \rho_{i-2}).}
\end{equation}
Recognizing the right-hand side as a discretization of the fourth derivative, we can take the continuum limit to recover the expected subdiffusive decay of the coarse-grained spin fluctuations~\cite{Feldmeier2020,Gromov2020}:
\begin{equation}\label{eq:sr_subdiffusive}
    \partial_t \rho(x,t) = -D_\text{sub} \partial_x^4 \rho(x,t),
\end{equation}
where $D_\text{sub}$ is a generalized diffusive constant that depends quadratically on the background spin density. This generalized Fick's law is a special case of a rank-2 continuity equation for dipole conservation, which in $d$ spatial dimensions has the form
\begin{equation}\label{eq:continuity_dip}
\partial_t \rho = - \partial_{a}\partial_{b} J_{ab},   
\end{equation}
where $J_{ab}$ is the current density of $x_a$-oriented dipoles moving in the $x_b$-direction~\cite{pretko2017subdimensional}. Restricting to $d=1$ and taking $J_{xx} = D_\text{sub} \partial_x^2 \rho$, we recover Eq.~\eqref{eq:sr_subdiffusive}.

For long-range couplings, however, the linearized master equation is more complicated, and there are different regimes, each of which corresponds to a different kind of hydrodynamic transport. Taking the Fourier transform of Eq.~\eqref{eq:master_dip_lin} we find
\begin{equation}\label{eq:master_dip_k}
        \partial_t \rho(k,t) = - A(k) \rho(k,t),
\end{equation}
where the momenta $k\in 2\pi \mathbb{Z}/L$ are restricted to the first Brillouin zone, $-\pi \leq k < \pi$, and the prefactor is given by 
\begin{equation}\label{eq:master_pre}
   A(k) = 4 \overline{\rho}(1-\overline{\rho}) \sum_{y =1}^L \sum_{n=1}^y  |J_{0,y,n}|^2 (1-\cos{ky})(1-\cos{kn}),
\end{equation} where we defined $y=i-j.$

For large system sizes, we can make progress by taking the continuum limit $L\rightarrow \infty$ while fixing $a=1,$ and hence replacing the sums with integrals. Concentrating on the long wavelength regime $k\ll 1/a$, which dominates the late-time dynamics, we will now focus on extracting the different scaling regimes of the master equation and their dependence on our choice of long-range couplings $J_{0,y,n}$. 
In general, the dominant long-wavelength behavior will be of the form
\begin{equation}\label{eq:general_kspace_diffeq}
\partial_t \rho(k,t) = -D k^{\eta} \rho(k,t),
\end{equation} 
where $D\sim K a^\eta$ is a generalized diffusive constant that depends on an effective dipole-exchange rate $K$, which is ultimately set by the underlying Hamiltonian, and the lattice constant $a$, whose appearance in the continuum limit is a consequence of the UV/IR mixing in systems with dipole symmetry~\cite{Gorantla2021, Gorantla2022}. In position space, this becomes a generalized Fick's law with a fractional Laplacian~\cite{Baggioli2021}:
\begin{equation}
    \label{eq:fractional_deriv}
    \partial_t \rho(x,t)  = - D (-\partial_x^2)^{\eta/2} \rho(x,t),
\end{equation}
and we can find the subdiffusion of Eq.~\eqref{eq:sr_subdiffusive} at $\eta=4$ or ordinary diffusion at $\eta=2$.

We can explicitly solve this differential equation in momentum space to find $\rho(k,t) = \rho(k,0) \exp\left[-D k^{\eta} t\right]$. Choosing an initial condition $\rho(x,0)\sim\delta(x)$ to model a localized packet of spin, the density in position space takes the form
\begin{equation}
    \label{eq:general_spatial_profile}
    \begin{split}
    \rho(x,t) &\sim \int \frac{dk}{2\pi} \exp[ikx - D t k^\eta] \\
    &= \frac{1}{(D t)^{1/\eta}} \mathcal{F}_\eta\left(\frac{\abs{x}}{(D t)^{1/\eta}}\right),
    \end{split}
\end{equation}
where $\mathcal{F}_\eta$ is a scaling function given by
\begin{equation}
    \label{eq:general_scaling_function}
    \mathcal{F}_\eta(u) \propto \int \frac{dk}{2\pi}\exp \left[iku - \abs{k}^{{\eta}}\right].
\end{equation}
This scaling function yields a symmetric generalized  distribution that reduces to a Gaussian for $\eta = 2$ or a Lorentzian for $\eta=1$.
It follows from Eq.~\eqref{eq:general_spatial_profile}, and the assumption of infinite temperature~\cite{Feldmeier2020}, that the equal-distance spin-spin correlator scales as $C(0,t)\sim t^{-1/\eta}$, indicating that the exponent of the dominant term in Eq.~\eqref{eq:master_pre} determines the dynamical exponent for spin transport, $\beta=1/\eta$.

Having described the theory for generic dipole-conserving couplings, we now consider some specific cases. We begin with the $\alpha_1,\alpha_2$-model of Eq.~\eqref{eq:alpha1alpha2}, for which the coupling takes the form
\begin{equation}\label{eq:alpha1alpha2_again}
    J_{0,y,n} = \frac{J_0}{|y|^{\alpha_1} |n|^{\alpha_2}}.
\end{equation} 
Before tackling this model in its full generality, it is instructive to restrict our attention to the $\alpha_2\rightarrow \infty$ limit, which describes long-range interactions between two dipoles of unit length. In this case, only the $n=1$ term of the second sum in Eq.~\eqref{eq:master_pre} survives, allowing us to expand the cosine and write the overall prefactor as
\begin{equation}
     \label{eq:dip_integral_special}
     A(k) \sim k^2 \int_1^\infty dy \frac{1-\cos{ky}}{|y|^{2\alpha_1}}.
\end{equation}
This integral is convergent as long as $\alpha_1 > 1/2$, and can be performed explicitly. However, we are mainly interested in the long-wavelength (small $k$) limit, which can be extracted by rewriting the integration region as $\int_1^\infty dy = \int_0^\infty dy - \int_0^1 dy$. Although extending the lower bound introduces spurious UV divergences for $\alpha_1>3/2$, they cancel between the two terms and can be safely ignored. In the bounded integral, we can use the $k\rightarrow 0$ limit to expand the cosine in the numerator and extract a factor of $k^2,$ while the unbounded integral scales like $k^{2\alpha_1-1}$. Combining these two results gives
\begin{equation}\label{eq:alpha1_scaling}
A(k) \sim C_1 k^{2\alpha_1+1} + C_2 k^4,
\end{equation}
where $C_1=-\Gamma(1-2\alpha_1)\sin(\pi\alpha_1),$ and $C_2 = 1/(2\alpha_1-3)$.  
The scaling of the spin fluctuations, $C(0,t)\sim t^{-\beta}$, is determined by which of these two terms dominates and depends on $\alpha_1.$ The first term dominates when $\alpha_1<3/2,$ and the second term dominates for $\alpha_1>3/2$. At $\alpha_1=3/2$, the two terms individually diverge, but their difference remains finite. In short, the dynamical exponent is given by $\beta^{-1} = \min(2\alpha_1+1,4)$. 
Furthermore, each coefficient $C_i$ is positive within its respective regime of dominance. 

In the regime $\alpha_1 < 1/2$, we note that Eq.~\eqref{eq:dip_integral_special} actually diverges. To cure this divergence we must replace the infinite upper bound on the integral by the IR cutoff given by the system size $L$. Then the integral simply yields an $\mathcal{O}(L^{1-2\alpha_1})$ constant that is canceled by the implicit Kac factor, leaving an overall scaling of $A(k)\sim k^2$, which implies diffusion.  

Collecting all of these results, we find the late-time behavior of the correlator $C(0,t)\sim t^{-\beta}$ as $\alpha_2\rightarrow \infty$:
\begin{equation}\label{eq:beta_dip_alpha1}
\beta^{(m=1)}(\alpha_1, \alpha_2=\infty) = 
\begin{cases}
1/4 & \alpha_1 \geq 3/2 \\
1/(2\alpha_1 + 1 ) & 1/2 < \alpha_1 < 3/2 \\
1/2 & \alpha_1 \leq 1/2.
\end{cases}
\end{equation}
When $\alpha_1 \ge 3/2$, \textit{dipole subdiffusion} prevails, meaning that $C(0,t)$ decays with the same dynamical exponent as in the short-range case. This is tied to the well-defined second moment of the probability distribution for dipole hopping when $\alpha_1 \ge 3/2$, resulting in a random walk of spins with an additional center of mass constraint.
In the intermediate regime $1/2 < \alpha_1 < 3/2$, the dynamical exponent is still subdiffusive, but smoothly interpolates between ordinary charge diffusion with $\beta=1/2$ and dipole subdiffusion with $\beta=1/4$. Remarkably, for even longer-ranged interactions with $\alpha\leq 1/2$, the correlator exhibits a stable diffusive regime, which persists even in the all-to-all connected case of $\alpha_1=0$. 

The pattern of dynamical exponents in Eq.~\eqref{eq:beta_dip_alpha1} is very similar to the analogous relaxation for a long-range charge conserving model (c.f. Eq.~\eqref{eq:beta_m0}). In fact, the scaling of the master equation for dipole conservation is identical except for an additional factor of $k^2$, which serves to slow down relaxation, and hence extend the hydrodynamic picture for the dipole case even in the presence of arbitrarily long-range couplings. The similarity can be formalized by the \textit{reciprocal rule}:
\begin{equation}\label{eq:reciprocal}
\frac{1}{\beta^{(1)}(\alpha_1, \alpha_2)} = \frac{1}{\beta^{(0)}(\alpha_1)} + \frac{1}{\beta^{(0)}(\alpha_2)},
\end{equation}
where for the moment we only consider the limit $\alpha_2\rightarrow \infty$ such that $\beta^{(0)}(\alpha_2) = 1/2$ (from Eq.~\eqref{eq:beta_m0}) and $\beta^{(0)}(\alpha_1) = 1/(2\alpha_1-1)$. The derivation of this relation is discussed in Appendix \ref{app:recip}, but essentially it is a direct consequence of the fact that the inverse dynamical exponents are additive in momentum space (see Eq.~\eqref{eq:general_kspace_diffeq}). Furthermore, the reciprocal rule encapsulates the different types of spin-spin interactions within a dipole-conserving gate like $S_i^+ S_{i+1}^- S_j^- S_{j+1}^+$. In particular, the second term in Eq.~\eqref{eq:reciprocal} arises from the local spin conservation inherent in independent spin exchange processes at sites $(i, i+1)$ and $(j, j+1)$, and reflects an underlying background of diffusive spin transport. However, these two spin exchanges are not really independent. The first term in Eq.~\eqref{eq:reciprocal} thus reflects the non-local constraint associated with the coordinated dipole-hopping between sites $i$ and $j$, which, depending on the degree of non-local interactions, can further slow the relaxation of spin. 

Conversely, we can also view the spin relaxation in long-range dipole-conserving systems as a speeding up of the behavior seen for short-range systems.
Although dipole conservation remains a valid global symmetry for any $\alpha_1$, only locally conserved charges govern hydrodynamic transport. For sufficiently long-ranged interactions, the dipole moment is no longer effectively conserved \emph{locally}. Instead, the exchange of two distant dipoles of unit length locally looks like just a spin-conserving process (see Fig.~\ref{fig:1}~(a)), resulting in precisely the diffusive dynamics that arise in cases with ordinary U(1) symmetry. In the intermediate regime between the $\beta=1/2$ and $\beta=1/4$, the long-range exponent $\alpha_1$ therefore tunes the ``locality'' of dipole conservation.

Now let us consider the general case where $\alpha_1, \alpha_2 < \infty.$ The momentum dependence of the prefactor $A(k)$ in the master equation is determined by the following integral (c.f. Eq.~\eqref{eq:master_pre}):
\begin{equation}\label{eq:dip_integral}    
A(k) \sim \int_1^\infty dy \int_1^y dz \frac{(1-\cos{ky})(1-\cos{kz})}{|y|^{2\alpha_1} |z|^{2\alpha_2}},
\end{equation}
which converges as long as $\alpha_1 > 1/2$ and $\alpha_1+\alpha_2>1$. We can handle this calculation like the $\alpha_2\to \infty$ case by recasting the integration into three different intervals:
(i) an unbounded interval in both variables, (ii) an unbounded interval just in $y$, and (iii) a bounded interval in both variables. In the case of a bounded domain, the $k\rightarrow 0$ limit allows us to expand the relevant cosine in the integrand and extract a factor of $k^2$, while no such expansion is possible for an unbounded domain. Once again, we treat the divergent cases at low $\alpha_i$ separately by re-inserting the IR cutoff and the Kac factor.

Collecting our results and extracting their $k$-dependence, we find that the overall prefactor in Eq.~\eqref{eq:master_dip_k} is now given by
\begin{equation}\label{eq:alpha12_scaling}
    A(k) = \begin{cases}
    A_1 k^{2(\alpha_1 + \alpha_2 -1)} + A_2 k^{2\alpha_1+1} + A_3 k^{4} & \alpha_1 \geq 1/2  \\
    B_1 k^{2\alpha_2 -1} + B_2 k^{2} & \alpha_1 < 1/2,
    \end{cases}
\end{equation}
where the coefficients $A_i$ depend on $\alpha_1$ and $\alpha_2$, and have similar expressions to those in Eq.~\eqref{eq:alpha1_scaling}. Given a choice of $\alpha_1$ and $\alpha_2$, the dominant term in Eq.~\eqref{eq:alpha12_scaling} for small $k$ determines the scaling of the spin fluctuations. 
It is instructive to rewrite the kernel above as
\begin{equation}
\label{eq:alpha12_scaling_2}
    A(k) = \begin{cases}
    A_1 k^{\tilde{\beta}^{-1}_1 + \tilde{\beta}^{-1}_2} + A_2 k^{\tilde{\beta}^{-1}_1+2} + A_3 k^{2+2} & \alpha_1 \geq 1/2  \\
    B_1 k^{\tilde{\beta}^{-1}_2} + B_2 k^{2} & \alpha_1 < 1/2,
    \end{cases}
\end{equation}
where we define the long-range charge-conserving exponents $\tilde{\beta}^{-1}_i = 2\alpha_i -1$. From Eq.~\eqref{eq:alpha12_scaling_2}, we can read off a generalized version of the reciprocal rule in Eq.~\eqref{eq:reciprocal}:
\begin{equation}\label{eq:beta_reciprocal_alpha_12}
    (\beta^{(1)})^{-1} = \begin{cases}
    \min (\tilde{\beta}^{-1}_1 + \tilde{\beta}^{-1}_2, \tilde{\beta}^{-1}_1+2, 2+2) & \alpha_1 \geq 1/2  \\
    \min(\tilde{\beta}^{-1}_2,2) & \alpha_1 < 1/2.
    \end{cases}
\end{equation}
As before, the dynamical exponent for a dipole-conserving system decomposes into a reciprocal sum of two charge-conserving exponents. Individually, these exponents can be either $\tilde{\beta}_i$ or $1/2$, respectively corresponding to long-range or short-range charge transport inside a dipole-conserving gate. Because $\alpha_2$ is effectively constrained by the range of dipole separation set by $\alpha_1$, there is an important asymmetry between the two exponents.
Indeed, when $\alpha_1<1/2$ the spin relaxation becomes independent of $\alpha_1$, and dipoles can hop arbitrarily long distances. Formally, this corresponds to taking $\tilde{\beta}_1^{-1}=0$ in the first line of Eq.~\eqref{eq:beta_reciprocal_alpha_12}.

We can reorganize Eq.~\eqref{eq:beta_reciprocal_alpha_12} in Table.~\ref{tab:recip_dipole}, which we will later generalize for the multipole conserving case in Sec.~\ref{sec:Analytic_quad}. The rows of the table represent the possible reciprocal sums, where the smallest row gives the dominant dynamical exponent, and the columns correspond to the different interactions contributing to those sums. 
Unlike the special case of $\alpha_2\rightarrow \infty$, the decay of the correlator can be faster than diffusive, and there are even regimes in which $\beta^{(m)}=0$. In such cases, the prefactor $A(k)$ is independent of $k$, and the charge relaxes exponentially fast, signaling a breakdown of the hydrodynamic description.
Interestingly, we see a hierarchical structure of the hydrodynamics in the $\alpha_1<1/2$ region, where we set $\tilde\beta_1^{-1}=0$ and obtain
\begin{equation}
(\beta^{(1)})^{-1} = \min(2\alpha_2-1,2) = (\beta^{(0)}(\alpha_2))^{-1}.
\end{equation}
In this regime, the interactions between dipoles are no longer local, and the spin relaxes as if only U(1) charge is locally conserved, i.e., the hydrodynamic description of dipole transport is reduced to a hydrodynamic description of charge transport.

\begin{table}[t!]
\begin{center}
\begin{tabular}{|c||P{1.2cm} P{1.2cm} P{1.2cm}|}
\cline{2-4}
\multicolumn{1}{c|}{} & \multicolumn{3}{c|}{$\tilde{\beta}^{-1}_n = \begin{cases}
    0 & \text{if } \alpha_{i\leq n} < 1/2 \\
    2\alpha_n - 1 & \text{otherwise}
\end{cases}$} \\
\hhline{-|===|}
\multirow{3}{*}{$(\beta^{(1)})^{-1}$} 
& \hspace{.5cm}  $\tilde{\beta}_1^{-1}$ & \hspace{.5cm}
 $+$ &  \hspace{.5cm} $\tilde{\beta}_2^{-1}$\\  
\cline{2-4}
& \hspace{.5cm}  $\tilde{\beta}_1^{-1}$ & \hspace{.5cm}
 $+$ &  \hspace{.5cm} $2$\\  
\cline{2-4}
& \hspace{.5cm}  $2$ & \hspace{.5cm}
 $+$ &  \hspace{.5cm} $2$\\  
\hline
\end{tabular}
\end{center}
\caption{Reciprocal rule for the dynamical exponent of a dipole-conserving system in terms of the dynamical exponents of a charge conserving system. The rows are reciprocal sums of long-range and short-range charge exponents. The smallest row sum gives the inverse dipole exponent $1/\beta^{(1)}$. }
\label{tab:recip_dipole}
\end{table}

A phase diagram depicting the various dynamical exponents of the spin-spin correlator for different values of $\alpha_1$ and $\alpha_2$ is shown in Fig.~\ref{fig:1}~(b). We first concentrate on the region $\alpha_2 \geq 3/2$, where only terms independent of $\alpha_2$ dominate the transport. This region of the phase diagram is qualitatively identical to the previously studied case of $\alpha_2\rightarrow \infty$, with a crossover between dipole subdiffusion and diffusion. On the other hand, the spin dynamics are much richer for $\alpha_2 < 3/2$. For instance, when lowering the long-range exponents along the line $\alpha_1 = \alpha_2$ we obtain regimes of subdiffusion, diffusion, superdiffusion (including ballistic transport, where $C(x,t)$ has a Lorentzian spatial profile), and ultimately the breakdown of hydrodynamics due to the exponential relaxation of spin. In all these regions, the specific value of the dynamical exponent can be extracted from the reciprocal rule in Eq.~\eqref{eq:beta_reciprocal_alpha_12}.

As shown in the second row of Fig.~\ref{fig:1}~(a), tuning $\alpha_1$ and $\alpha_2$ establishes a hierarchy of \emph{local} conservation laws for gates that \emph{globally} conserve dipole moment. For $\alpha_1, \alpha_2 \rightarrow \infty$, the gates are entirely local, and the dipole moment is locally conserved, giving rise to dipole subdiffusion. Tuning $\alpha_1$ to finite values interpolates between local dipole conservation and local charge conservation, and subsequently tuning $\alpha_2$ to finite values then interpolates between local charge conservation and no locally conserved charge at all. Throughout this hierarchy, the dynamical exponent $\beta^{(1)}(\alpha_1, \alpha_2)$ increases in accordance with the extent of the local conservation laws. The same pattern can also be seen in Table~\ref{tab:recip_dipole}: when $\alpha_1 < 1/2$, the first column of reciprocal exponents vanishes, and the spin transport is described by a single long-range parameter $\alpha_2$ that tunes between diffusion and exponential relaxation. This is identical to the hydrodynamics of charge-conserving systems~\cite{Schuckert2020}, meaning that the dipole symmetry no longer affects local spin transport. 

Let us momentarily turn away from the $\alpha_1, \alpha_2$-model to consider the experimentally motivated long-range coupling arising from trapped ions in a tilted potential, Eq.~\eqref{eq:JTilt}. In this case, the leading order rate in the master equation is given by:
\begin{equation}\label{eq:JTilt_rate}
W_{0,y,n}\propto\frac{1}{|y|^{2(\gamma+1)}|n|^{2(\gamma+1)} |y-n|^{2\gamma}},
\end{equation}
where $\gamma \geq 0$ is the long-range exponent of the underlying trapped-ion XY model~\cite{Morong2021}. 
Using the linearized master equation~\eqref{eq:master_dip_lin} we obtain:
\begin{equation}\label{eq:dip_tilt_integral}
    A(k) \sim \int_1^\infty dy \int_1^y dz \frac{(1-\cos{ky})(1-\cos{kz})}{|y|^{2\gamma+2} |z|^{2\gamma+2} |y-z|^{2\gamma}},
\end{equation}
which converges for all allowed values of $\gamma$. Unlike the previous case, there are only two regimes as encoded in
\begin{equation}\label{eq:JTilt_scaling}
    A(k) = \tilde{A}_1 k^{6\gamma +2} + \tilde{A}_2 k^4,
\end{equation}
where the first term becomes dominant for $\gamma<1/3$. In practice, this implies that dipole subdiffusion is extraordinarily stable to long-range Ising interactions between ions in a tilted potential. For very long-range systems with $\gamma<1/3$, the dynamical exponent interpolates between dipole subdiffusion and diffusion. Even in the case of all-to-all couplings in the original Hamiltonian ($\gamma=0$), the spin transport remains diffusive, a reflection of the particular constraints arising from the perturbative expansion of the original model. 
\subsection{Long-range hydrodynamics with quadrupole and higher multipole symmetry}\label{sec:Analytic_quad}
Now let us consider systems where, in addition to the charge and dipole moment, the quadrupole moment is also conserved. We expect the transport of spin to slow down even further compared to the charge and dipole cases. Here, we briefly consider the master equation for the long-range, quadrupole-conserving couplings of the $\alpha_1, \alpha_2, \alpha_3$ model defined in Eqs. (\ref{eq:GeneralQuadrupoleConservingHam}-\ref{eq:Q_coefficient}). We recall that the coupling is represented by
\begin{equation}
    Q_{i,j,n_1,n_2} = \frac{Q_0}{|i-j|^{\alpha_1} |n_1|^{\alpha_2} |n_2|^{\alpha_2}},
\end{equation}
where $\alpha_1$ sets the distance between quadrupoles, $\alpha_2$ sets the distance between their constituent dipoles, and $\alpha_3$ sets the distance between their respective constituent spins.

In the limit $\alpha_2, \alpha_3 \rightarrow \infty,$ the linearized master equation for the spin density $\rho_i$ takes the following form:
\begin{equation}
    \label{eq:master_quad_lin}
    \frac{d \rho_i(t)}{dt} = -{\overline{\rho}}^3(1-\overline{\rho})^3 \sum_{i\neq j} |Q_{i,j}|^2 \left[ \Delta^{(4)}_x \rho_j - \Delta^{(4)}_x \rho_i \right],
\end{equation}
where $\Delta^{(s)}_x$ is the lattice discretization of the $s$th spatial derivative. Repeating the same arguments used for the dipole-conserving case (c.f. Eq.~\eqref{eq:master_dip_lin}), we can extract the long-wavelength scaling: 
\begin{equation}\label{eq:alpha1_scaling_quadrupole}
A(k) \sim C_1 k^{2\alpha_1+3} + C_2 k^6,
\end{equation}
where the only difference from Eq.~\eqref{eq:alpha1_scaling} is an additional overall factor of $k^2$ coming from the higher derivatives in the master equation.
As $\alpha_1$ is lowered, the spin transport now exhibits a crossover between subdiffusion with $\beta=1/6$, characteristic of short-range quadrupole-conserving systems, and dipole subdiffusion with $\beta=1/4$.

Extending this result to non-infinite $\alpha_2, \alpha_3$ is straightforward, one has only to expand the lattice derivatives above and extend their range by $n_1$ and $n_2$. The result is an integral expression similar to  Eq.~\eqref{eq:dip_integral} that dictates the scaling of the ensuing hydrodynamic differential equation. Once again, the dynamical exponent can be essentially captured using the reciprocal rule:
\begin{equation}\label{eq:reciprocal_quad}
\frac{1}{\beta^{(2)}(\alpha_1, \alpha_2, \alpha_3)} = \frac{1}{\beta^{(0)}(\alpha_1)} + \frac{1}{\beta^{(0)}(\alpha_2)} + \frac{1}{\beta^{(0)}(\alpha_3)},
\end{equation} where, like before, care must be taken to treat the interdependence of different exponents (see discussion below Eq. \eqref{eq:beta_mpole} and Appendix \ref{app:recip}) in regimes where one of them dominates the spin transport.  The resulting phases of spin relaxation are qualitatively similar to the ones shown for the dipole case in Fig.~\ref{fig:1}~(b), though with a greater variety of subdiffusive exponents and crossovers. Furthermore, there is a parallel hierarchy of local conservation laws tuned by the long-range couplings, ranging from local conservation of all globally conserved moments down to  no local conservation of any of the moments, as shown in the third row of Fig.~\ref{fig:1}~(a).

For completeness, we also briefly sketch the linearized master equation and dynamical exponents for $m$th moment-conserving systems in one dimension. We consider only $\alpha_1, \ldots, \alpha_{m+1}$-models with a hierarchical separation of spins into their constituent $m$-pole hopping terms, like the dipole and quadrupole couplings we have thus far examined. For the case of ultralocal $m$-poles undergoing long-range hopping processes set by a single exponent $\alpha_1$, we find the linearized master equation
\begin{equation}
    \label{eq:master_lin_all}
    \hspace{-4mm}
    \frac{d \rho_i(t)}{dt} = -\left[{\overline{\rho}}(1-\overline{\rho})\right]^{2^m-1} \sum_{i\neq j} \frac{1}{|i-j|^{2\alpha_1}} \left[ \Delta^{(2m)}_x \rho_j - \Delta^{(2m)}_x \rho_i \right],
\end{equation}
which leads to 
\begin{equation}\label{eq:beta_dip_alpha1_m}
\beta^{(m)}(\alpha_1, \alpha_i=\infty) = 
\begin{cases}
1/(2m+2)& \alpha_1 \geq 3/2 \\
1/(2\alpha_1 + 2m -1) & 1/2 < \alpha_1 < 3/2 \\
1/2m & \alpha_1 \leq 1/2.
\end{cases}
\end{equation}
It is clear that for $\alpha_1\ge 3/2$, we obtain ``$m$-pole subdiffusion'' like the short-range case, while for $\alpha_1\le 1/2$ the system relaxes as if only the $(m-1)$th moments are locally conserved, yielding $\beta = 1/(2(m-1)+2)=1/2m$.  

To obtain the master equation for the full $\alpha_1, \ldots, \alpha_{m+1}$-model, where the $m$-poles can be longer, we need to expand the finite differences in the sum to arbitrarily long range, i.e., making the replacement $\rho_{i+1}, \rho_{i+2}, \ldots \rightarrow \rho_{i+n_1}, \rho_{i+n_1+n_2}, \ldots$ in Eq.~\eqref{eq:master_lin_all}. 
Taking the Fourier transform and the continuum limit, we find long-wavelength hydrodynamics governed by:
\begin{equation}\label{eq:master_all_k}
\begin{split}
\partial_t \rho(k,t) \sim \rho(k,t) \sum_{r=0}^{m+1} A_r k^{\gamma_r}, \\
\gamma_r  = 2r + \sum_{i\leq m-r+1} (2\alpha_i -1) 
\end{split}
\end{equation} where $A_r$ are coefficients that depend on the values of the $\{\alpha_i\}$, and Eq.~\eqref{eq:master_all_k} holds for $\alpha_i\geq 1/2$. 

As we discuss in more detail in App.~\ref{app:recip}, the dynamical exponent for $m$-pole conserving hydrodynamics is hierarchical and can be related to the dynamical exponents for charge hydrodynamics by a generalized reciprocal rule. More precisely, we find
\begin{equation}
    \label{eq:beta_mpole}
    \frac{1}{\beta^{(m)}(\{\alpha_i\})} = \min_{r\in[0,m+1]} \gamma_r,
\end{equation}
 where the $\gamma_r$ are given in Eq.~\eqref{eq:master_all_k}, and are sums of a set of inverse charge exponents $1/\beta^{(0)}(\alpha_i)$.  These exponents characterize the various spin-spin couplings within an $m$-pole conserving gate. Their values  range from diffusive ($1/\beta^{(0)} =2$) to superdiffusive ($1/\beta^{(0)} =2\alpha_i - 1$) based on their associated long-range strength $\alpha_i$.
If any of the $\gamma_r$ are negative, then the hydrodynamic approximation breaks down and charge relaxes exponentially. 

Although the $\gamma_r$ in Eq. \eqref{eq:master_all_k} are only well-defined for $\alpha_i \geq 1/2$,  we can extend our reciprocal rule to $\alpha_i<1/2$ by keeping track of which long-range exponents are dominant over others (see App.~\ref{app:recip}). For example, when $\alpha_1 \rightarrow \infty$ our models become short-ranged, and no other exponents $\alpha_{i>1}$ affect the spin relaxation.  Similarly, when $\alpha_n \rightarrow \infty$ no subsequent exponents $\alpha_{i>n}$ matter. When the dominant exponents are small, i.e., $\alpha_{j\leq n} < 1/2$ for some integer $n$, then the reciprocal rule must be amended by the replacement $(2\alpha_j-1) \rightarrow 0$, as we showed for the dipole-conserving case in Sec.~\ref{sec:Analytic_dip}.  This adjustment reflects the fact that the relevant dynamical exponent for charges, $\beta^{(0)}(\alpha_{j\leq n})$, diverges in Eq.~\eqref{eq:beta_m0}.

We summarize the full reciprocal rule in Table~\ref{tab:recip}. Much like the dipole-conserving case (Table~\ref{tab:recip_dipole}), the reciprocal of the $m$-pole dynamical exponent is given by the smallest row sum in the table (since smaller $\gamma_r$ will give a more dominant exponent). We have discussed how the hierarchical structure of the hydrodynamics arises  because we can decompose $m$-pole conserving dynamics into a combination of lower-moment conserving events, and ultimately to its  constituent charge conserving events. Indeed, it is this phenomenology that is precisely codified in the reciprocal rule and Table \ref{tab:recip}.  For example, taking $\alpha_1 < 1/2$, and therefore an ultra-long-range coupling between $m$-poles, $\tilde \beta_1^{-1} = 0.$ In this parameter regime there is no \emph{local} $m$-pole conserving dynamics, though we can still have local $(m-1)$-pole conservation (and so forth). Since this parameter regime effectively erases the first column of the table, the reciprocal rule then reduces to the rule for the case of local $(m-1)$-pole conservation,
 \begin{equation}
 (\beta^{(m)})^{-1} = \min_{r\in [0,m]} \gamma_r  = (\beta^{(m-1)})^{-1},\quad \alpha_1<1/2.
 \end{equation}
Recursively, we obtain the full hierarchical structure
 \begin{equation}
 (\beta^{(m)})^{-1} = \min_{r\in [0,m+1-r]} \gamma_r  = (\beta^{(m-r)})^{-1},\quad \alpha_{i\le r}<1/2.
 \end{equation} In this sense, the reciprocal rule not only gives the dynamical exponent, but also encodes the hierarchy of generalized diffusion arising from the hierarchy of long-range couplings between multipoles.

\begin{table}[t!]
\centering
\begin{tabular}{|c||c c c c c c c c c|}
\cline{2-10}
\multicolumn{1}{c|}{} & \multicolumn{9}{c|}{$\tilde{\beta}^{-1}_n = \begin{cases}
    0 & \text{if } \alpha_{i\leq n} < 1/2 \\
    2\alpha_n - 1 & \text{otherwise}
\end{cases}$} \\
\hhline{-|=========|}
$\gamma_0$ & $\tilde{\beta}^{-1}_1$ & + & $\tilde{\beta}^{-1}_2$ & + & $\cdots$ & + & $\tilde{\beta}_{m}^{-1}$ & + & $\tilde{\beta}_{m+1}^{-1}$ \\
\hline
$\gamma_1$ & $\tilde{\beta}^{-1}_1$ & + & $\tilde{\beta}^{-1}_2$ & + & $\cdots$ & + & $\tilde{\beta}_{m}^{-1}$ & + & $2$ \\
\hline
$\vdots$ & $\vdots$ & & $\vdots$ & & &  & $\vdots$ & & $\vdots$
\\
\hline
$\gamma_m$ & $\tilde{\beta}^{-1}_1$ & + & $2$ & + & $\cdots$ & + & $2$ &+& $2$\\
\hline
$\gamma_{m+1}$ & $2$ & + & $2$ & + & $\cdots$ & + & $2$ & + & $2$\\ 
\hline
\end{tabular}
\caption{Reciprocal rule for the dynamical exponent of an $m$-pole conserving system in terms of the dynamical exponents of a charge conserving system, the latter of which we define as $\beta^{(m=0)}(\alpha_i) \equiv \tilde{\beta}_i$ for simplicity. The rows are reciprocal sums of long-range and short-range ($\beta^{-1}=2$) charge relaxation exponents. The inverse $m$-pole exponent $1/\beta^{(m)}$ is given by the smallest row sum $\gamma_r$.}
\label{tab:recip}
\end{table}

\subsection{Beyond one dimension}\label{sec:higher_dimensions}
We have thus far restricted our analyses to one spatial dimension, but the master equation approach can be readily generalized to an arbitrary number of dimensions. A generic dipole-conserving process involves the exchange of two anti-parallel, equal-length dipoles. For example, the $\alpha_1,\alpha_2$-model in $d$-spatial dimensions becomes
\begin{equation}
    \label{eq:dipole_ddim}
    H = \sum^{'}_{\bm{r}_1,\bm{r}_2,\bm{n} \in \mathbb{R}^d} \frac{J_0}{|\bm{r}_1-\bm{r}_2|^{\alpha_1} |\bm{n}|^{\alpha_2} } \left(S^+_{\bm{r}_1} S^-_{\bm{r}_1+\bm{n}} S^-_{\bm{r}_2} S^+_{\bm{r}_2+\bm{n}} + \text{h.c.}\right),
\end{equation}
where $\bm{r}_i$ are the positions of two length-$\bm{n}$ dipoles that are swapped in a dipole-hopping process, and the prime over the sum indicates that we take $|\bm{n}| < |\bm{r}_1-\bm{r}_2|$ to avoid overcounting. 

The linearized master equation for such dipole-exchange processes can again be written in the form of Eq.~\eqref{eq:master_dip_k} with coefficient
\begin{equation}\label{master_pre_ddim}
A(k) \sim \int_{r>1} d^d \bm{r} \int_{1<r'<r} d^d\bm{r}' \frac{(1-e^{i k r \cos{\theta}})(1-e^{i k r' \cos{\theta'}})}{r^{2\alpha_1}r'^{2\alpha_2}},
\end{equation}
where $k=|\bm{k}|$, and the integral converges if $\alpha_1>d/2$ and $\alpha_1+\alpha_2>d$. Expanding at small $k$ to extract the long-wavelength behavior, we find that the hydrodynamic equation scales like
\begin{equation}\label{eq:diffeq_ddim}
\partial_t g(k) = \left[A^{(d)}_1 k^{2(\alpha_1 + \alpha_2 - d)} + A^{(d)}_2 k^{2\alpha_1 + 2 - d} + A^{(d)}_3 k^{4}  \right]g(k),
\end{equation}
where the $A^{(d)}_i$ are $\alpha_i$-dependent coefficients that enter as generalized diffusion constants. It is clear from the scaling in \eqref{eq:diffeq_ddim} that the dipole subdiffusive term $k^4$ becomes less dominant for large $d$, implying that long-range couplings have a more dramatic effect on dipole-conserving dynamics in higher dimensions. Given a dominant term $\partial_t \rho \sim k^\eta \rho$, we can follow Eq.~\eqref{eq:general_spatial_profile} to find the (isotropic) real space profile of the resulting spin fluctuations:
\begin{equation}
    \label{eq:general_spatial_ddim}
    \rho(\bm{x},t) \sim \frac{1}{(D t)^{d/\eta}} \mathcal{F}^{(d)}_\eta\left(\frac{\abs{\bm{x}}}{(D t)^{1/\eta}}\right),
\end{equation}
where the scaling function $\mathcal{F}^{(d)}_\eta$ is given by
\begin{equation}
    \label{general_scaling_ddim}
    \mathcal{F}^{(d)}_\eta(u) \propto \int dk \, k^{d-1} e^{-k^\eta} \int_0^\pi d\theta \sin^{d-2}({\theta}) \, e^{i ku \cos{\theta}}.
\end{equation}

\begin{figure}[t!]
    \includegraphics[width=1.\columnwidth]{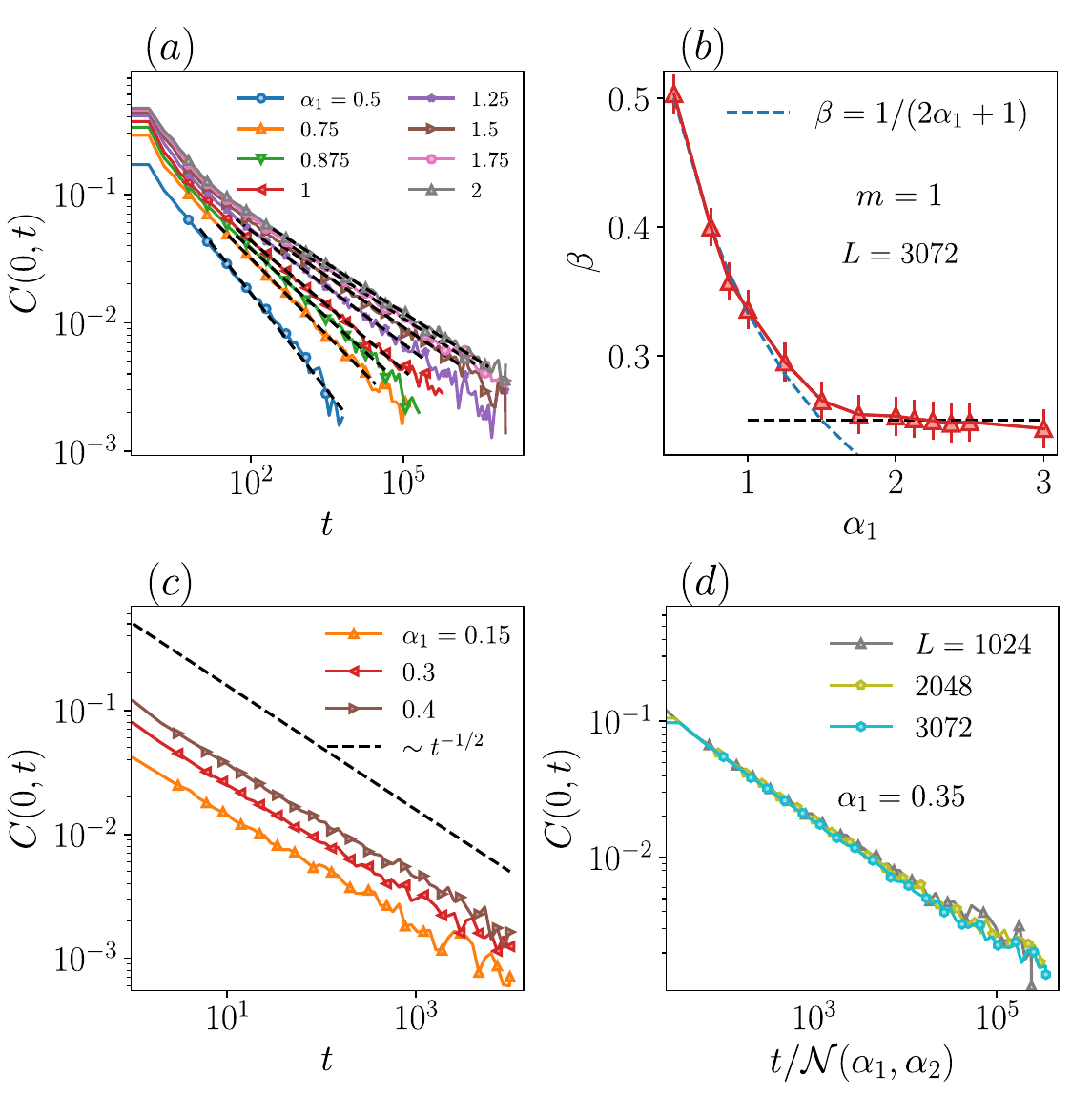}
    \caption{(a) Return probability $C(x=0,t)$ for the dipole-conserving case with $\alpha_2\rightarrow \infty$ for several values of $\alpha_1$. (b) Dynamical exponent $\beta$, where $C(0,t)\sim t^{-\beta}$, as a function of $\alpha_1$. For $\alpha_1>3/2$, we have $\beta\sim 1/4$ as in short-range models, and the dashed line represents the analytical prediction. (c) Behavior of the return probability for small values of $\alpha_1$. The black dashed line represents diffusive relaxation and serves as a guide for the eye. (d) $C(0,t)$ for several system sizes $L$ with $\alpha_1=0.35$. The time is rescaled with the Kac factor.}
    \label{fig:Full_Dipole_alpha_1}
\end{figure}
The full phase diagram in $d$-dimensions is similar to the one-dimensional case shown in Fig.~\ref{fig:1}~(b), but with subdiffusion giving way to diffusion and superdiffusion at higher values of $\alpha_1$ and $\alpha_2$. For example, moving along the line $\alpha_1=\alpha_2\equiv \alpha$, we find two hydrodynamic phases:
\begin{equation}
    \label{eq:2d_phases}
    \beta^{(m=1)}_{(d)}(\alpha) \sim \begin{cases}
        d/(4\alpha - 2d) & d/2 < \alpha < 1+ d/2  \\
        d/4 & \alpha \geq 1+d/2.
    \end{cases}
\end{equation}
Below $\alpha = d/2$, hydrodynamics breaks down and spin relaxes exponentially fast. We can also consider the limit $\alpha_2\rightarrow \infty$, in which case the crossover between $\beta=d/4$ and faster subdiffusion is still at $\alpha_1=1+d/2$, but for $\alpha \leq d/2$ the system remains stably diffusive due to the effective local spin conservation. The extension of these results to higher-multipole conserving systems described by additional $\alpha_i$ can be carried out straightforwardly.

It is clear from Eq.~\eqref{eq:2d_phases} that in high enough dimensions even a weak long-range coupling is enough to destroy dipolar subdiffusion. Intuitively, this is because lattices in higher dimensions have more spins available to participate in dipole exchanges, weakening the impact of the kinematic dipole constraint. Indeed, since lattice connectivity increases with increasing dimension, it is easier for a dipole to find a partner to exchange with inside some fixed radius. From an alternative perspective, decreasing the long-range exponent $\alpha$ also effectively augments lattice connectivity, and $\alpha$ heuristically behaves as an inverse dimension in Eqs.~(\ref{eq:diffeq_ddim}-\ref{eq:2d_phases}).
This relationship between the long-range exponent $\alpha$ and the spatial dimension $d$ is a generic feature of long-ranged systems, which in many cases exhibit the same localization behavior~\cite{Mirlin1996} and phase transitions~\cite{Fisher1972, Defenu2021} as short-ranged systems in an effective dimension $d_\text{eff} \sim d/\alpha$. As we have seen, the spin relaxation in our systems is yet another property where the long-range strength $\alpha$ can play the role of an inverse dimension. 

\subsection{Numerical results} \label{sec:Numeric}
After having established the behavior of $C(x,t)$ and determined its dynamical exponent $\beta$ as a function of the Hamiltonian parameters $\{\alpha_i\}$ and the degree of the highest conserved moment $m$, we confirm our results by performing stochastic dynamics using cellular automata, as described in Section~\ref{sec:Methods}. For concreteness, we fix $S=1$ in the dipole-conserving case ($m=1$) and $S=3$ in the quadrupole case ($m=2$). 

We begin our numerical analysis with the $\alpha_1, \alpha_2$ model, Eq~\eqref{eq:alpha1alpha2}, by considering $\alpha_2\rightarrow \infty$. In this limit, dipoles of length one can hop by a distance $r$ with a probability $\sim r^{-2\alpha_1}$. In Section~\ref{sec:Analytic_dip}, we analytically established that for $\alpha_1>3/2$, we find   subdiffusive dynamics identical to the short-ranged model with an exponent of $\beta = 1/4$; at long times we expect $C(k,t)\sim e^{-D k^4t}$. For $1/2<\alpha_1<3/2$, the dynamics is also subdiffusive, but with an $\alpha$-dependent exponent $\beta = 1/(2\alpha_1+1)$, while for smaller $\alpha_1$ a stable diffusive phase with $\beta = 1/2$ is present. 
\begin{figure}[t!]
\includegraphics[width=1.\columnwidth]{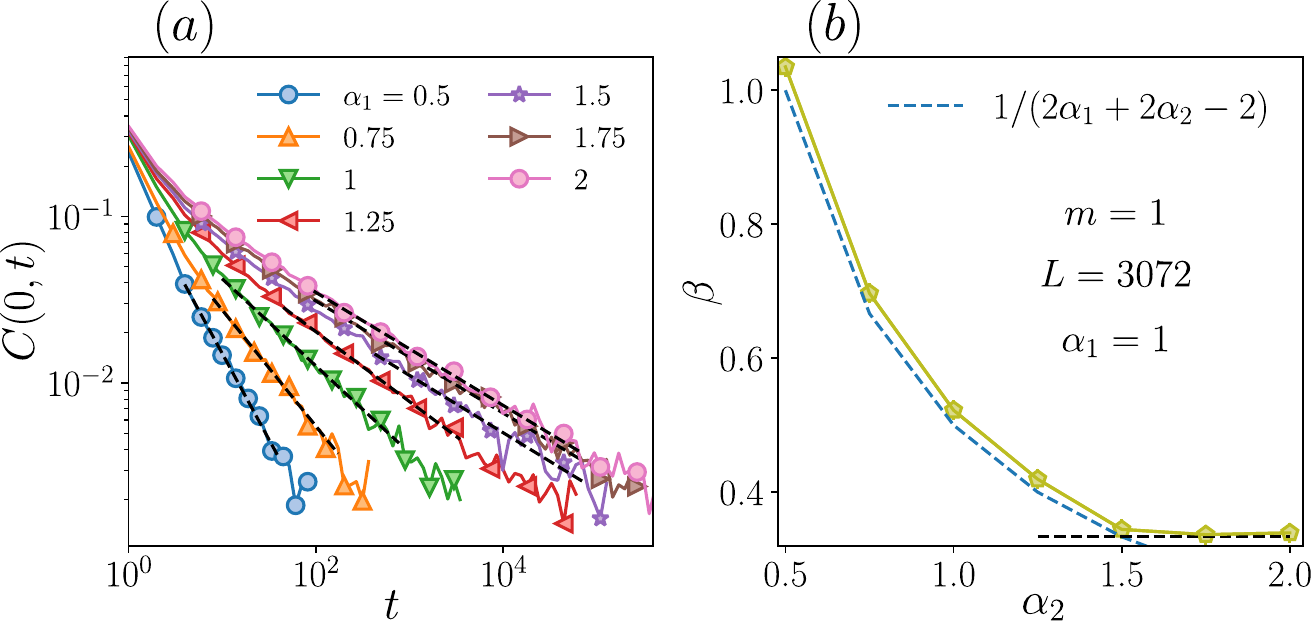}
    \caption{(a) $C(0,t)$ for the dipole-conserving case with $\alpha_1=1$ for several values of $\alpha_2$. (b) The dynamical exponent $\beta$ extracted from $C(0,t)$ in (a). For $\alpha_2>3/2$, we have $\beta \sim 1/3$ (horizontal black dashed line), while the blue dashed line represents the analytical prediction for $\alpha_2<3/2$.}
    \label{fig:Two_exponents}
\end{figure}

We show numerical evidence to corroborate our analytic results in Fig. \ref{fig:Full_Dipole_alpha_1}. Explicitly, Fig.~\ref{fig:Full_Dipole_alpha_1}~(a) shows $C(0,t)$ as a function of $t$ for several orders of magnitude up to $t \approx 10^7$ and for several values of $\alpha_1$, with fixed system size $L=3072$. As expected, the relaxation of $C(0,t)$ is algebraic, faster for smaller values of $\alpha_1$, and slower for larger values. We extrapolate $\beta$ by performing a fit of $C(0,t)$ at lat times, as shown in Fig.~\ref{fig:Full_Dipole_alpha_1}~(b). In agreement with our theoretical prediction, $\beta\sim 1/4$ for $\alpha_1>3/2$. At $\alpha_1=3/2$, a logarithmic correction accounts for the slight deviation from the expected value~\cite{Schuckert2020}.
For intermediate values, the numerically extrapolated slopes follow the analytical prediction $\beta \sim 1/(2\alpha_1+1)$ (see dashed line in Fig.~\ref{fig:Full_Dipole_alpha_1}~(b)). 

Next, we show the behavior of $C(0,t)$ for small $\alpha_1$ in Figs.~\ref{fig:Full_Dipole_alpha_1}~(c)-(d). Here the dynamics is universal for $\alpha_1<1/2$, and $C(0,t)\sim t^{-1/2}$ (see Fig.~\ref{fig:Full_Dipole_alpha_1}~(c)) . Notably, $C(0,t)\sim t^{-1/2}$ also occurs for systems that conserve locally charge/magnetization. Such a connection is expected, since for small values of $\alpha_1$, the dipoles that are created and removed by the dipole hopping term tend to be very far away from each other. Hence, only the magnetization is conserved over small length scales (see the second row of the table in Fig.~\ref{fig:1}~(a)). 
It is important to point out that in this regime the time scale depends on the system size because the energy is superextensive. In order to obtain $L$-independent dynamics, we renormalize the time in Fig.~\ref{fig:Full_Dipole_alpha_1}~(d) by the relevant Kac factor, $\mathcal{N}_{\alpha_1,\alpha_2}= \sqrt{\sum_{n_1=1}^L {n_1}^{-2\alpha_1} \sum_{n_2=1}^{n_1} n_2^{-2\alpha_2}}$, which in the limit $\alpha_2\rightarrow \infty$ is of order $L^{(1-2\alpha_1)/2}$. 
\begin{figure}[t!]
    \includegraphics[width=1.0\columnwidth]{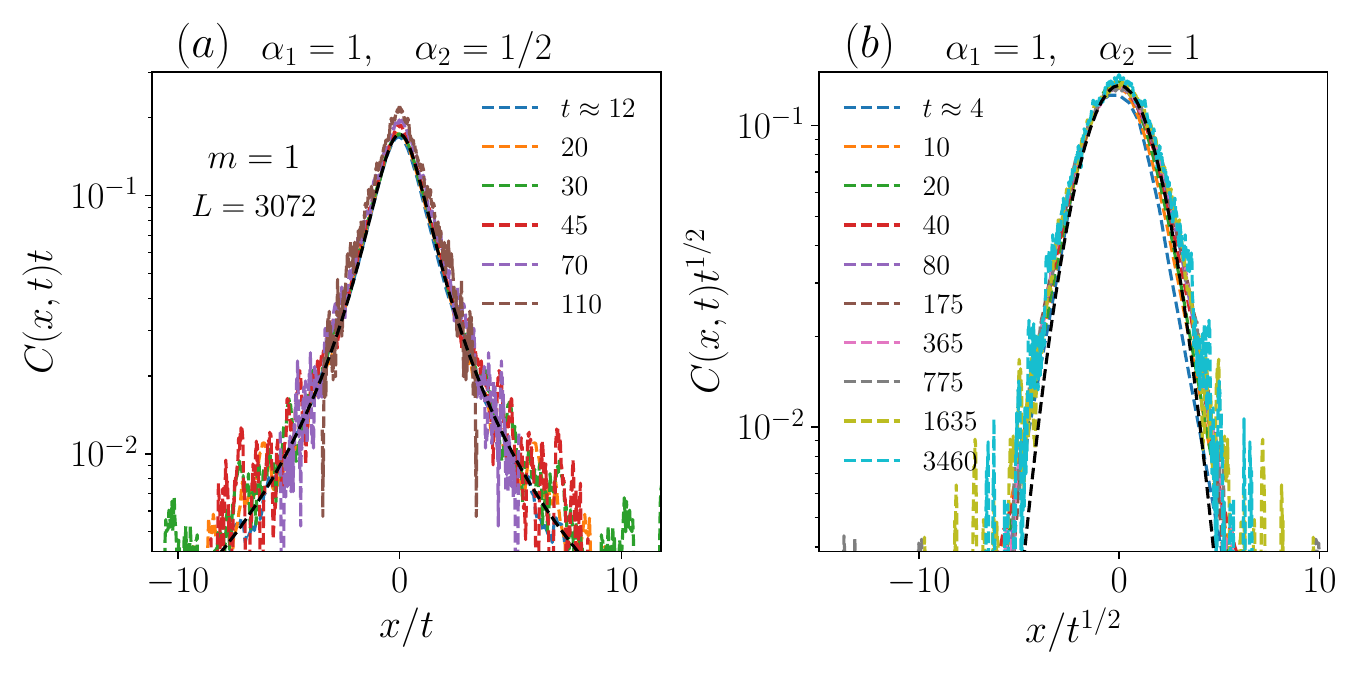}
    \caption{The full spatial profile $C(x,t)$ for several times. Each panel represents a different $\alpha_2$, with fixed $\alpha_1=1$ and $m=1$. The axes have been rescaled to collapse the curves. The black dashed lines in (a) and (b) are Lorentzian ($\alpha_2=1/2$) and Gaussian ($\alpha_2=1$) profiles, respectively.}
    \label{fig:Profile_two_exponents}
\end{figure}

Returning to the general case, we also test our results regarding the behavior of $\beta$ in Eq.~\eqref{eq:alpha12_scaling} for $\alpha_2<\infty$. In particular, in Fig.~\ref{fig:Two_exponents}~(a)-(b) we show the results of tuning $\alpha_2$ while keeping $\alpha_1=1$ fixed. We find that the numerics confirm our prediction based on the reciprocal rule $1/\beta = 1/\beta^{(0)}(\alpha_1) + 1/\beta^{(0)}(\alpha_2) = 1/{2\alpha_2}$ for $\alpha_2<3/2$ (see dashed line in Fig.~\ref{fig:Two_exponents}~(b)). Thus, there are different regimes: subdiffusion for $\alpha_2>1$, superdiffusion for $\alpha_2<1$ (Lévy flights), and ballistic behavior at $\alpha_2=1/2$. For larger $\alpha_2>3/2$, the relaxation is subdiffusive with $\beta = 1/3$, highlighted by the dashed line in Fig.~\ref{fig:Two_exponents}~(b). 

For the sake of completeness, we analyze the full shape of $C(x,t)$ for two interesting cases, $(\alpha_1,\alpha_2)=(1,1/2)$ and $(1,1)$, in Figs.~\ref{fig:Profile_two_exponents}~(a) and (b), respectively. 
In Fig.~\ref{fig:Profile_two_exponents}~(a)-(b), we show $C(x,t)$ as a function of $x$ for several chosen target times. Crucially, we collapse our observables by rescaling both the $x$-axis and the magnitude of $C(x,t)$ by $t^{-\beta}$. In Sec.~\ref{sec:Analytic_dip}, we showed that rescaled two-point correlator function takes a universal shape given by Eq.~\eqref{eq:general_scaling_function}.
In particular, for $\alpha_2=1/2$, we are in the Lévy flight regime, the transport is ballistic, and $\mathcal{F}_{1}(y)$ in Eq.~\eqref{eq:general_scaling_function} takes the form of a Lorentzian. In comparison, for $\alpha_2=1$ the system is diffusive, and the universal function is a Gaussian (see the black dashed lines in Fig.~\ref{fig:Profile_two_exponents}). 

\begin{figure}[t!]
\includegraphics[width=1.\columnwidth]{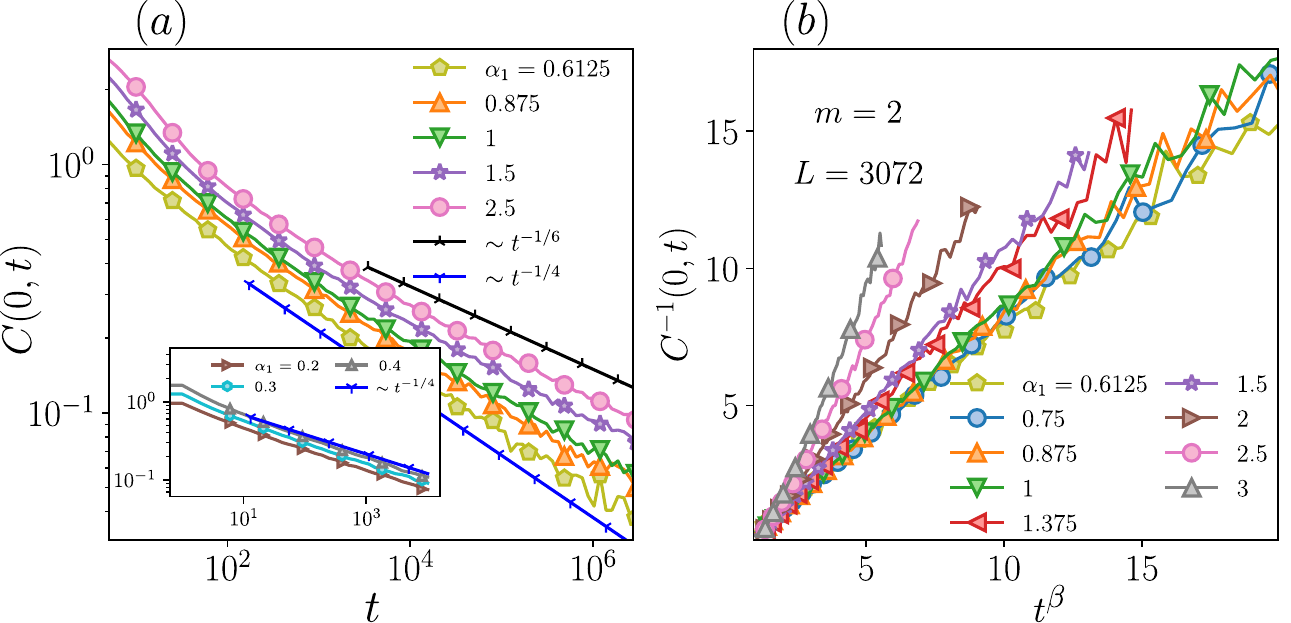}
    \caption{(a) $C(0,t)$ for the quadrupole-conserving case with $\alpha_2\rightarrow \infty$ and several values of $\alpha_1$. The dark blue and black guidelines correspond to the limiting cases: dipole subdiffusion, $\sim t^{-1/4}$, for small $\alpha_1$, and quadrupole subdiffusive, $\sim t^{-1/6}$, for large $\alpha_1$. The inset illustrates $C(0,t)\sim t^{-1/4}$ for small $\alpha_1<1/2$. (b) We parameterize the time to show $C(0,t)\sim t^{-\beta}$ with $\beta$ given by the theoretical prediction.}
    \label{fig:Quadrupole}
\end{figure}

Having confirmed our conjectures on the behavior of $\beta$ for the dipole case, we now turn to the quadrupole case. The conservation of quadrupole moment, assuming that both dipole moment and charge conservation are also conserved, imposes even stronger constraints and leads to slower dynamics than the dipole case. For example, short-range models are characterized by anomalous diffusion with $\beta = 1/6$. Such slow dynamics require many time steps in numerical simulations to extract the dynamical exponent $\beta$, which negatively impacts the numerical performance. To this end, we focus on the case $S=2$ to increase the system's mobility level, and we consider the limiting case $\alpha_{2},\alpha_{3}\rightarrow \infty$ in Eq.~\eqref{eq:Q_coefficient}. In this limit, compact (length-three) quadrupoles at positions $i$ and $j$ experience algebraically decaying interactions $1/{|i-j|^{\alpha_1}}$. Analytically, we previously found that (i) for $\alpha_1>3/2$ the dynamical exponent is $\beta = 1/6$, (ii) in the regime $1/2<\alpha_1 <3/2$ the exponent grows as $\beta = 1/(2\alpha_1+2 m -1)=1/(2\alpha_1 +3)$, and (iii) for $\alpha_1<1/2$ the dynamics is universal with $\beta\sim 1/4$, and is dominated by the locally conserved dipole moment (see the third row of the table in Fig.~\ref{fig:1}~(a)). 
In Fig.~\ref{fig:Quadrupole}~(a), we show $C(x=0,t)$ for several $\alpha_1$ at a fixed system size $L=3072$. The dark blue and black guide lines in Fig.~\ref{fig:Quadrupole}~(a) indicate the two limiting cases, $C(0,t)\sim t^{-1/4}$ and $\sim t^{-1/6}$, for small and large values of $\alpha_1$, respectively. The inset in Fig.~\ref{fig:Quadrupole}~(a) provides evidence that the dynamics is universal for $\alpha_1<1/2$ where $C(0,t)\sim t^{-1/4}$, as discussed. To support our analytical predictions and to avoid fitting,  we plot $C^{-1}(0,t)$ as a function of $t^\beta$. As one can observe in Fig~\ref{fig:Quadrupole}~(b), $C^{-1}(0,t)$ is, to good approximation (after early time transient behavior), directly proportional to $t^{\beta}$, providing further evidence in support of our predictions in Sec.~\ref{sec:Analytic_quad}.

\begin{figure}[t!]
    \includegraphics[width=1.\columnwidth]{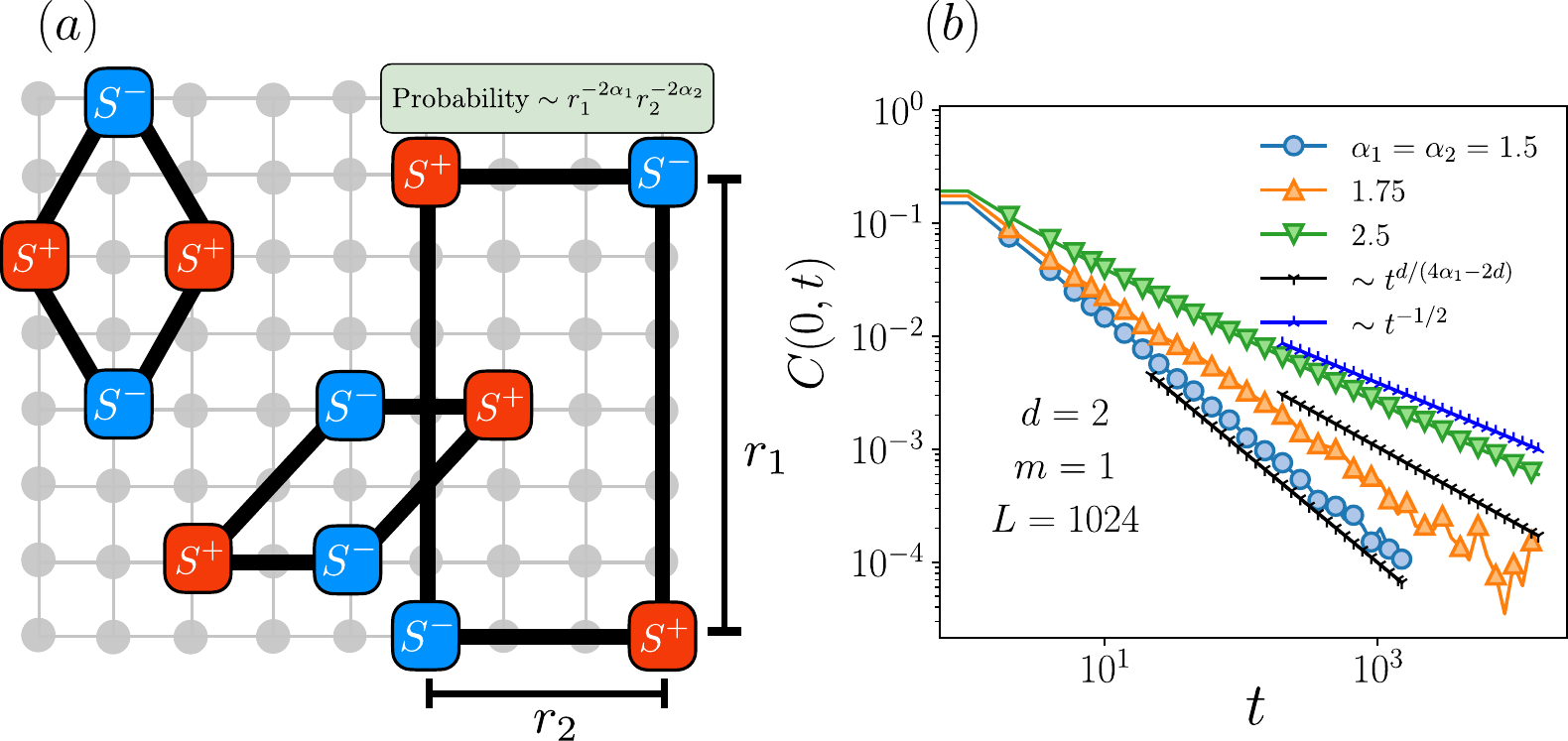}
    \caption{(a) Schematic illustration of two-dimensional dipole-conserving gates applied within a single timestep, where the probability to apply the gate decays algebraically with dipole separation, (b) Two-point correlator $C(0,t)$ calculated with two-dimensional cellular automata gates for $\alpha_1=\alpha_2$. The blue line represents dipole subdiffusion with $\beta=d/4$, while the black lines indicate faster relaxation with $\beta$ in Eq.~\eqref{eq:2d_phases}~(higher and lower lines correspond to $\alpha=1.75$ and $\alpha=1.5$, respectively). }
    \label{fig:Two_dimensions}
\end{figure}

Now, we turn to the question of how the dynamical exponent changes in dimensions higher than one ($d\ge 2$). In Sec.~\ref{sec:higher_dimensions}, we derived an expression for $\beta$ in arbitrary dimension $d$ for the dipole-conserving case. To test our prediction, we consider the two-dimensional case, and take the interaction between spins to algebraically decay with exponent $\alpha_1=\alpha_2$ in Eq.~\eqref{eq:2d_phases}. We fix $S=1$ and consider the most generic gates that act on four spins and conserve $P^{(1)}$, see Fig.~\ref{fig:Two_dimensions}~(a). In Fig.~\ref{fig:Two_dimensions}~(b), we show $C(0,t)$ as a function of time for a few values of $\alpha_1$, calculated in a two-dimensional square lattice of size $1024\times 1024$. As expected, we observe an algebraic relaxation $C(0,t)\sim t^{-\beta}$ and the exponents $\beta(\alpha)$ are in relatively in good agreement with our theoretical prediction in Eq.~\eqref{eq:2d_phases} (see black and blue lines in Fig.~\ref{fig:Two_dimensions}~(b)).   

\subsection{Away from zero-magnetization}
So far, we have always computed the correlator by averaging over random initial configurations, effectively probing the the zero-charge sector $P^{(m)}=0$, i.e., the largest sector in the Hilbert space. In this section, we extend our results beyond the zero charge sector for a representative example. We focus on the dipole-conserving case with $S=1$ and $\alpha_2\rightarrow \infty$ in Eq.~\eqref{eq:alpha1alpha2}. For simplicity, we define the variable $n_i=S_i^z+1 \in \{0,1,2\}$. In terms of this new variable, the system is equivalent to particles hopping on a lattice with the constraint that no more than two particles can occupy a single site. 
\begin{figure}[t!]
\includegraphics[width=1.\columnwidth]{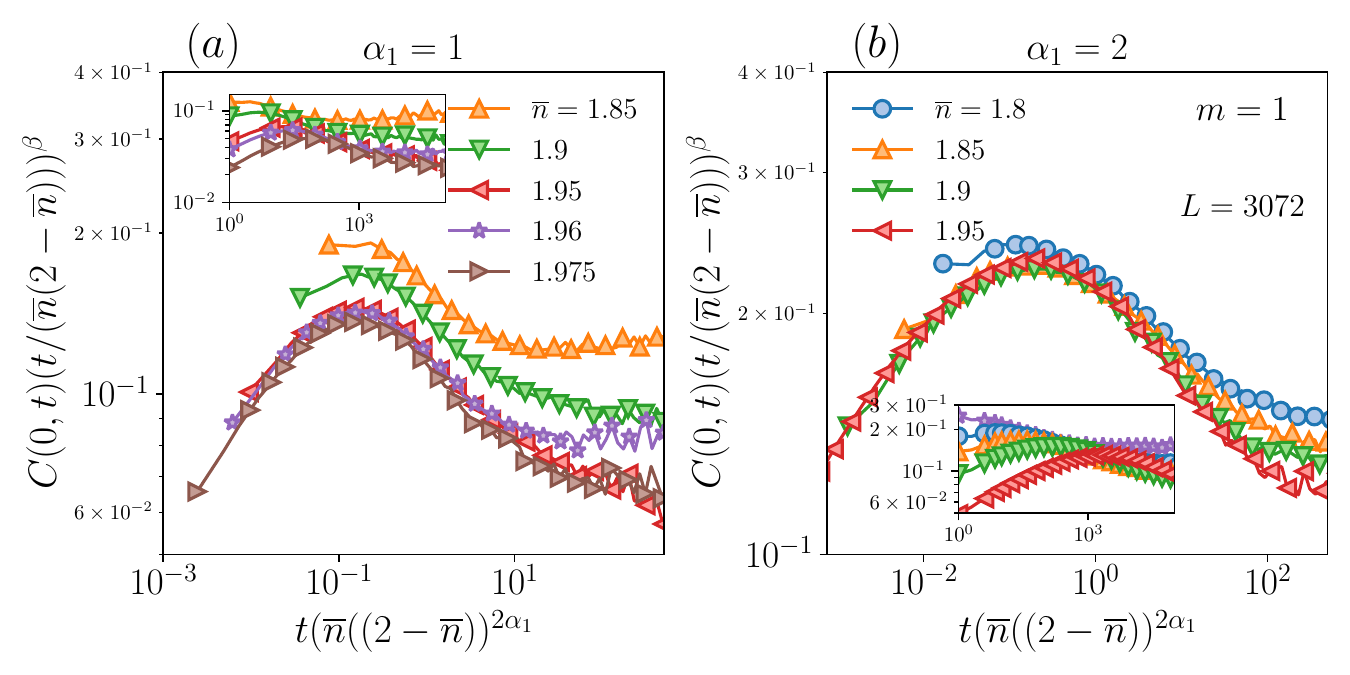}
    \caption{The rescaled two-point correlator $C(0,t)$ for several values of the density $\overline{\rho} = (\langle S^z_i \rangle + S)/2$, taking overall spin $S=1$, for (a) $\alpha_1=1$ and (b) $\alpha_1=2$. The time has been rescaled by $t^{\star} \sim 1/{(\overline{\rho}(1-\overline{\rho} ))^{2\alpha_1}}$, and $C(0,t)$ has been rescaled by its long-time behavior $\sim (\overline{\rho}(1-\overline{\rho})/t)^\beta$ (which is asymptotically exact in the limit of high/low density). The insets in (a) and (b) depict $C(0,t)t^{\beta}$  without rescaling time by $t^{\star}$.}
    \label{fig:Finite_density}
\end{figure}
As demonstrated in Refs.~\onlinecite{Morningstar2020,Pozderac23}, dipole-conserving systems with short-range ($k$-local) interactions exhibit a freezing transition driven by particle density that separates a weakly fragmented Hilbert space phase from a strongly fragmented Hilbert space phase. 
Moving away from half-filling, $\overline{\rho} = \langle n_i \rangle/2 = 1/2$, frozen bubbles of high or low charge density emerge and prevent particles from moving through the system. In the spin language, these bubbles are contiguous regions of spins that are ``stuck'' because of dipole conservation.  Heuristically, the freezing transition occurs when the typical length of these frozen bubbles, $\ell_\text{frozen}$, exceeds the range of interactions. 
Because our interactions have power-law decay, there is always the possibility (perhaps with low probability) that a spin inside a frozen bubble can flip by coupling to a very distant spin. Eventually, this will always happen, but there can be a long time scale before the freezing transition is washed out by the long-range interactions.
Such emergent time scales are similar to prethermal regimes, and slow, nearly frozen dynamics are observed for a long time before relaxation begins, i.e., when $C(0,t)\sim t^{-\beta}$. 

To understand how this prethermal time scale arises, it is useful to consider the limiting case in which the particle density is almost maximal (minimal), and frozen bubbles are regions of contiguous sites with the same spin. In this limit, $\overline{\rho}\rightarrow 1$~($\overline{\rho}\rightarrow 0$), and the mean length of a frozen bubble is $\ell_\text{frozen}\sim 1/[{\overline{\rho}(1-\overline{\rho})]}$.  We note that $W(\ell_\text{frozen}) \sim \ell^{-2\alpha_1}_{frozen}$ is the probability of moving active particles across a frozen region of length $\ell_\text{frozen}$ set by Fermi's golden rule, since $\ell_\text{frozen}$ is the shortest distance over which a dipole exchange process can occur. Hence, for time scales given by $t^{\star}(\alpha_1, \overline{\rho})\sim W^{-1}(\ell_\text{frozen})$, we expect a prethermal regime characterized by slow propagation.  However, at asymptotically large times, $t\gg t^{\star}$, spins can undergo multiple long-range jumps across frozen regions, and we anticipate that the dynamical exponent $\beta$ becomes independent of the particle density, which only enters as a prefactor in the diffusive constant, much like the usual Einstein relation for Brownian motion. Indeed, from our master equation analysis in Eq.~\eqref{eq:master_dip_lin}, we can deduce that $C(0,t)\sim (\overline{\rho}(1-\overline{\rho})/t)^{\beta}$. 
\begin{figure}[t!]
\includegraphics[width=1.\columnwidth]{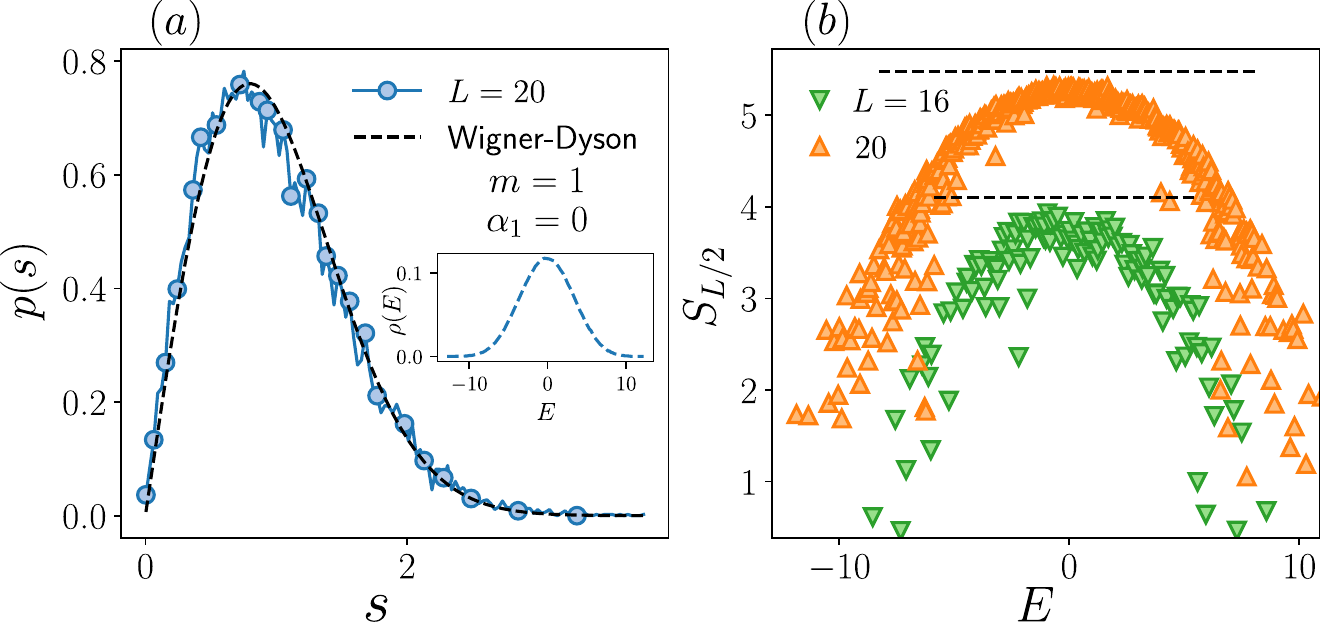}
\caption{(a) Level spacing probability distribution for energies located in the middle of the spectrum, $E\approx 0$, of Eq.~\eqref{eq:H_Quantum_dipole}. The black dashed line represents the Wigner surmise of chaotic systems, while the inset in (a) illustrates the many-body density of states $\rho(E)$. (b) Half-partition entanglement entropy $S_{L/2}$ as a function of energy for $L=16$ and $20$, for one random configuration of $H_Q$. The dashed lines indicate the mean value of $S_{L/2}$ for random states having $P^{(0)}= P^{(1)}=0$.}
    \label{fig: Dipole_quantum_eig}
\end{figure}
To test the emergence of the time scale $t^{\star}$, we perform stochastic automaton simulations for several values of the average particle density $\overline{\rho}$, as shown in Fig.~\ref{fig:Finite_density}. To analyze the behavior of $C(0,t)$, we first rescale it by its asymptotic behavior $(\overline{\rho}(1-\overline{\rho})/t)^{\beta}$, so that at long times the curves approach a plateau. We identify the time scale $t^{\star}$ as the point where $C(0,t^\star)/\left[\overline{\rho}(1-\overline{\rho})/t^\star\right]^\beta$ reaches its maximum value, as shown in Fig.~\ref{fig:Finite_density} and its insets. After that, we rescale the time by our prediction for $t^{\star} \sim \ell_\text{frozen}^{2\alpha_1}$ to align the maxima. In agreement with our heuristic scaling for $t^{\star}$, in Fig.~\ref{fig:Finite_density} we see a good collapse of the curves, and especially an alignment of their maxima. Thus, for $t<t^\star$, $C(0,t)$ decays slower than $t^{-\beta}$, while for larger times, we start to approach relaxation with $C(0,t)\sim t^{-\beta}$. Importantly, this time scale diverges in the limit of trivially frozen dynamics $\overline{\rho}\rightarrow 0,1$. Therefore, for any fixed $\alpha_1$, we can tune $\overline{\rho}$ to obtain parametrically large time scales in which the dynamics are almost frozen.

\subsection{Numerical results for quantum dipole-conserving model}
After investigating the dynamics of long-range systems with conserved higher-moments through classical cellular automaton simulations, we now focus on examining the robustness of our findings in the truly quantum scenario. Investigating the complete quantum evolution presents significant obstacles because the Hilbert space grows exponentially as the system size increases. Consequently, we are restricted to relatively small system sizes and timescales, making it challenging to access the anticipated hydrodynamic regime~\cite{Lux_14}.

To reduce possible finite size effects and increase the maximal extent $L$ of our one-dimensional system, we focus on only the dipole-conserving case ($m=1$), spin-$1/2,$ and $\alpha_2\rightarrow \infty$ in Eq.~\eqref{eq:alpha1alpha2}. Furthermore, to increase the level of ergodicity for finite system sizes, we slightly modify our Hamiltonian
\begin{equation}
    H_Q = \frac{1}{\mathcal{N}_{\alpha_1}}\sum_{i, j>i+1} \left( \gamma_{i,j}  \frac{S^+_i S^-_{i+1} S^-_j S^+_{j+1}}{|i-j-1|^{\alpha_1}}+ \text{h.c.}\right) + \sum_i h_i S^z_i,
\label{eq:H_Quantum_dipole}{}
\end{equation}
where $\gamma_{i,j}=\gamma_{j,i}$ are random-sign random variables, and $h_i~\in~[-1,1]$ are small random fields. We use open boundary conditions and take the Kac factor $\mathcal{N}_{\alpha_1} = \sqrt{\sum_{n_1} n_1^{-2\alpha_1}}$, which, as mentioned above, makes the energy extensive in system size instead of superextensive. 
The Hamiltonian conserves both charges (magnetization) and dipole moment, and we focus our analysis on the largest Hilbert space sector given by $\sum_j S_j^z = \sum_j j S^z_j = 0$. 
\begin{figure}[t]
    \includegraphics[width=1.\columnwidth]{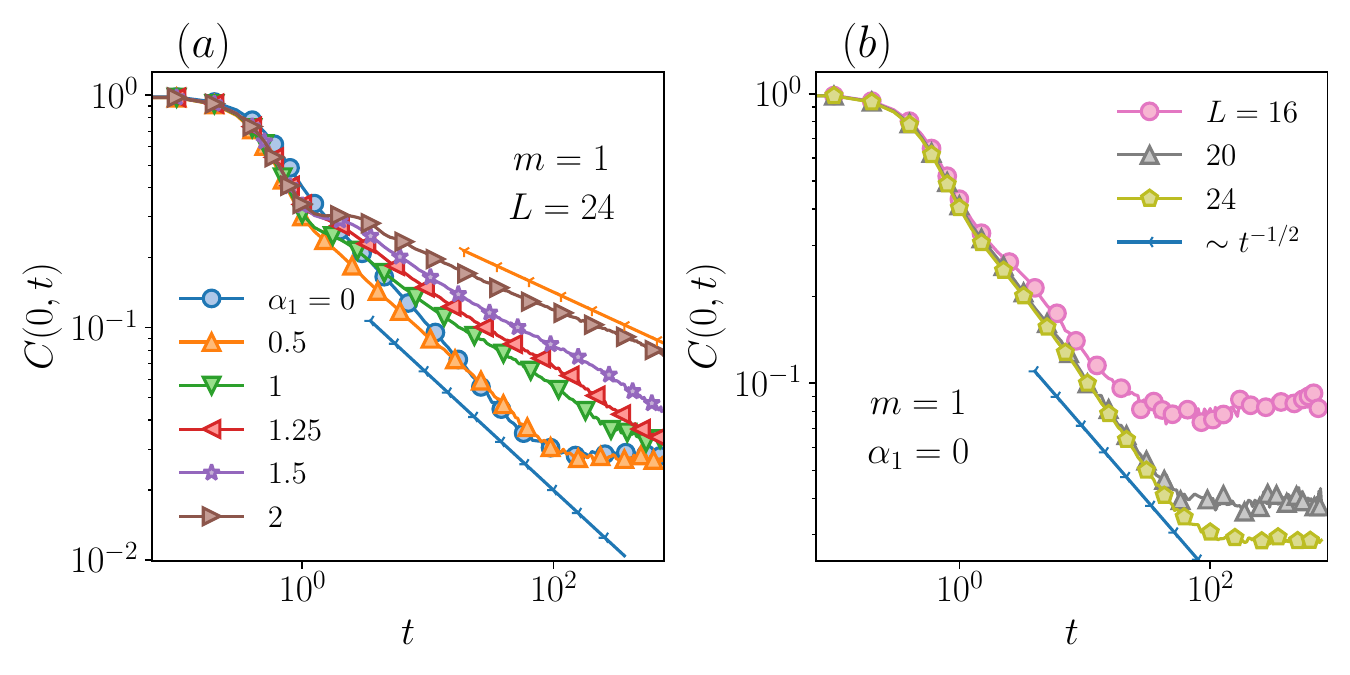}
    \caption{(a) Quantum calculation of $C(0,t)$ as a function of time for $L=24$ and for several values of $\alpha_1$. The straight blue and orange lines correspond to the two limiting cases: diffusive, $\sim t^{-1/2}$ (blue), and dipole subdiffusive, $\sim t^{-1/4}$ (orange), respectively. (b) $C(0,t)$ for fixed $\alpha_1$ and several system sizes $L$. The straight blue line corresponds to diffusive behavior.}
    \label{fig: Dipole_quantum_Dyna}
\end{figure}

For completeness, we briefly demonstrate the ergodicity of our Hamiltonian for our system sizes. We first examine the probability distribution of nearby energy levels as a probe. In chaotic systems, adjacent energy levels exhibit level repulsion. For our model at $\alpha_1=0,$ we find that the probability distribution of $s = (E_{n+1} - E_n) / \langle E_{n+1} - E_n \rangle$ assumes the form of the Wigner surmise: $p(s) = \frac{\pi s}{2} e^{-\pi s^2/4}$, as shown in Fig.~\ref{fig: Dipole_quantum_eig}~(a) for energy levels in the middle of the spectrum of Eq.~\eqref{eq:H_Quantum_dipole}. Levels obeying this distribution exhibit level repulsion. Furthermore, we investigate the behavior of the half-system von-Neumann entanglement entropy, $S_{L/2} = -Tr[\rho_{L/2}\log{\rho_{L/2}}]$, where $\rho_{L/2}$ is the reduced density matrix for an eigenstate of $H_Q$. As shown in Fig.~\ref{fig: Dipole_quantum_eig}~(b), $S_{L/2}$ exhibits the typical rainbow shape as a function of energy $E$ characteristic of chaotic quantum systems. Furthermore, at infinite temperature (in the middle of the spectrum with $E\approx 0$), $S_{L/2}$ approaches its ergodic values, shown as horizontal black dashed lines in Figs.~\ref{fig: Dipole_quantum_eig}~(b). To summarize, we can assume that for the system sizes we consider, $H_Q$ in Eq.~\eqref{eq:H_Quantum_dipole} shows a good degree of chaoticity. 

Now, we turn to the out-of-equilibrium dynamics of $H_Q$ and inspect the two-point correlator $C(0,t)$. For the quantum case, the average $\langle \cdots \rangle$ in Eq.~\eqref{eq:correlator} should be interpreted as the normalized trace over the full Hilbert space. We perform the quantum time evolution using Chebyshev integration techniques and perform the Hilbert space trace stochastically using the concept of quantum typicality~\cite{Wiesse_06}. The results for $C(0,t)$ are shown in Fig~\ref{fig: Dipole_quantum_Dyna}. Despite being limited to a maximum system size of $L=24$, we observe an algebraic relaxation $C(0,t)\sim t^{-\beta}$, where the dynamical exponent $\beta,$ to some approximation, respects the bounds $\beta \sim 1/2$ for small values $\alpha_1 \lessapprox 1/2$, and tends to $\beta \sim 1/4$ for larger ones. 

\section{Discussion }~\label{sec:Discussion}

In this work, we examined the interplay between multipole conservation and long-range, algebraically decaying couplings. Our primary interest was the long-time dynamics, which are believed to be dominated by the underlying conserved charge moments, giving rise to an emergent hydrodynamic description. Using this approach, we described a hierarchical sequence of models that globally conserve up to $m$th multipole moments but locally perhaps conserve only a subset of them. Within the hydrodynamic framework, the nature of the long-time transport reflects conservation laws on a local scale, so tuning the range of the interactions generates a variety of subdiffusive, diffusive, and superdiffusive regimes of spin/charge relaxation.

Despite the breadth of transport phenomena observed for different choices of the algebraically decaying interactions, in all cases, there is a stable region of subdiffusion that exactly matches the dynamics of short-range models that conserve the $m$th-moment, i.e., $C(0,t)\sim t^{-1/(2m+2)}$. In practice, this implies that experimental efforts to detect signatures of fractonic behavior through anomalously slow thermalization can succeed even when the underlying platforms, like cold atoms and trapped ions, exhibit power-law (Coulomb) interactions. In fact, we showed that the spin relaxation of a long-range XY model in a strong tilted potential is captured by a long-range dipole-conserving model that exhibits quite stable subdiffusion, with a dynamical exponent of $\beta=1/4$ that persists up to inverse cube root interactions $V(r)\sim r^{-1/3}.$ Furthermore, ordinary diffusion in such systems arises only in the all-to-all connected limit. 

Fundamentally, subdiffusion is stable because when $\alpha\geq3/2$ the probability distribution for dipole exchange has a finite second moment, and therefore the individual charges undergo an ordinary random walk (with a dipole constraint) like in the short-range case. In passing, we note that this also implies that subdiffusion with $\beta=1/4$ is even more stable than the heuristic scaling argument in Ref.~\onlinecite{Gromov2020} suggests, since the long-range hopping probability mimics a short-ranged one as soon as its second moment is finite~\footnote{The authors of Ref.~\onlinecite{Gromov2020} argue that when $\alpha>d/2+2$, long-range interactions do not destroy dipole subdiffusion, but (depending on which $\alpha_1, \alpha_2$ we consider) the true bound for subdiffusion can be much lower in $\alpha_i$ (see Fig.~\ref{fig:1}~(b))}. 

Throughout this work, we have considered long-time, infinite temperature regimes where the quantum dynamics are well approximated by classical hydrodynamics. It would be interesting to understand how  quantum measures of complexity, like entanglement, propagate in dipole-conserving systems. In a similar vein, it would also be useful to compare the spread of different types of multipole operators in the presence of long-range interactions. Even within the classical hydrodynamic picture we have not treated nonlinear regimes of the master equation that describe stronger fluctuations in spin density. In the short-range case, these effects are described by a nonlinear sub-diffusive equation for the density of the form $\partial_t \rho \sim  - \partial_x^4 \rho - (\partial_x^2 \rho)^2 $, and the additional term can give rise to a range of localization and scaling phenomena~\cite{Han2023} that would be interesting to tune with a long-range exponent.

Additionally, our work has focused on dynamics at infinite temperature. Still, it may also be extended to finite temperature, where a new energy-time scale emerges that can make relaxation even slower. 
On a different front, long-range interactions may also significantly alter the equilibrium phases of multipole-conserving systems. In particular, because tuning the long-range exponents $\alpha$ can change the effective spatial dimension, a richer pattern of spontaneous dipole symmetry breaking~\cite{Stahl2022, Lake2022a, Lake2022b, Glorioso2023, Armas2023, Jain2023} may be possible in low-dimensional experimental platforms with algebraically decaying interactions.
Lastly, we point out that studying long-range systems with  subsystem symmetries is an exciting direction for future research.

\section{Acknowledgments}
We thank P. Sala and J. H. Han for insightful discussions. GDT would like to thank J.  Feldmeier, M. Knap, and F. Pollmann for an earlier collaboration on a similar topic and for many insightful discussions.
J.G., J.M.M., and T.L.H. thank ARO MURI W911NF2020166 for support. 
J.M.M. is also supported by the National Science Foundation Graduate Research Fellowship Program under Grant No. DGE - 1746047. 
G.D.T. acknowledges support from the EPiQS Program of the Gordon and Betty Moore Foundation.
\\

\noindent {\it Note added.}---During the preparation of this manuscript, we became aware of independent works by A. Morningstar, N. O'Dea, and J. Richter~\cite{MorningstarRichter2023} and O. Ogunnaike, J. Feldmeier, and J. Y. Lee~\cite{Ogunnaike2023}
that also consider long-range interactions in dipole-conserving systems. Both works appeared in the same arXiv posting, and we thank the authors for coordinating submission.

\bibliographystyle{apsrev4-2}
\bibliography{dipole.bib}

\appendix
\section{Effective Hamiltonian for a long-range XY model in a tilted potential}\label{app:EffHamiltonian}
This appendix will present the effective Hamiltonian for a long-range XY model in a tilted potential. Our starting point is the 1D Hamiltonian 
\begin{equation}
    H = \sum_{j,k} t_{i,j} S^+_i S^-_{j} + \text{h.c.} + \sum_k k F  S^z_j,
\label{eq:RydbergWithTilt}\end{equation}
where the $t_{i,j}$ terms are long range XY-couplings between the spins at sites $i$ and $j$, and $F$ sets the strength of the tiled potential. In the limit where $t_{i,j} \rightarrow 0$, the spectrum of the system divides into sectors with different dipole moments. For finite $t_{i,j}<<F$, we can construct the effective dipole-conserving Hamiltonian that acts on of these sectors. 

Using a Schrieffer-Wolff (SW) transformation, the effective Hamiltonian up to second order in $t_{i,j}/F$ is 
\begin{equation}\begin{split}
     H_{\text{eff}} = &\sum_{i<j} \frac{1}{4F}  S^z_i \frac{(t_{i,j})^2}{i-j} \\& + \sum_{i<j<k} \frac{1}{2F^2}  S^z_i S^z_j \frac{t_{i,j} t_{j,k} t_{k,i}}{(i-k)(j-k)}\\
     & + \sum_{i<j<k<l} \sum_{n} J_{i,j,k,l} S^+_i S^-_{j} S^-_{k} S^+_{l} + \text{h.c.}
\label{eq:EffectiveHam}\end{split}\end{equation}
where 
\begin{equation}\begin{split}
    J_{i,j,k,l} = -\frac{1}{4F^2} &\Big ( \frac{t_{i,j} t_{i,k} t_{i,l}}{(i-j)(i-k)} + \frac{t_{i,l} t_{j,l} t_{k,l}}{(l-j)(l-k)} \\
    &  - \frac{t_{i,j} t_{j,k} t_{j,l}}{(i-j)(j-l)} - \frac{t_{i,k} t_{j,k} t_{k,l}}{(i-k)(k-l)} \Big ) 
\end{split}\end{equation}
If we take the XY-coupling $t_{j,k}$ to have the power law form, $t_{j,k} = t_0 \frac{1}{|j-k|^\gamma}$, then $J_{i,j,k,l}$ takes the form given in Eq.~\eqref{eq:JTilt} with $J_0 = -\frac{t_0^3}{4F^2}$. 

\section{Derivation of the reciprocal rule for the multipole relaxation exponent}\label{app:recip}
Using the same linearized master equation approach as for the dipole case (c.f Sec.~\ref{sec:Analytic_dip}), the long-time decay of the charge density in an $m$-pole conserving system obeys
\begin{equation}
\partial_t \rho(k,t) =  A(k) \rho(k,t).
\end{equation}
Here the prefactor $A(k)$ in the long-wavelength limit determines the dynamical exponent $\beta^{(m)}$, and is given by 
\begin{equation}
\label{eq:general_prefactor}
A(k) \sim \int_1^\infty dx_1 \frac{1-\cos{kx_1}}{x_1^{2\alpha_1}}\prod_{i=1}^{m}\int_1^{x_i} dx_{i+1} \frac{1-\cos{kx_{i+1}}}{x_2^{2\alpha_{i+1}}}.
\end{equation}

The nested ``Russian doll'' form of the integration region, which formally resembles the time-ordered expression for the unitary time evolution operator in the Dyson series, makes the convergence of the rightmost integrals depend on that of the previous ones. In other words, the dynamics are dominated by the exponents $\alpha_i$ with the lowest index $i$.

First, we consider the regime $\alpha_i \ge 1/2$ so that all the integrals converge. The dominant term will give the dynamical exponent in the long-wavelength limit of Eq.~\eqref{eq:general_prefactor}. To this end, each integral can be split into a noncompact domain and a compact domain, $\int_1^\infty dx_i = \int_0^\infty dx_i - \int_0^1 dx_i$. Along any compact domain, we can expand the cosine in the $k\rightarrow 0$ limit and the resulting integral scales like $k^2$. We cannot expand the cosine for noncompact domains; instead, we extract an overall factor of $k^{2\alpha_i-1}$ for each integration variable $x_i$ by rescaling. 

Performing this expansion from left to right in the integrals, we get what may appear to be $2^{m+1}$ different terms. However, if an integral over $x_i$ has a compact domain, all the following integrals over $x_{j>i}$ also have a compact domain since the leftmost integration variable bounds them. Because of this asymmetry, we find the following pattern of overall scaling:
\begin{equation}\label{eq:nested_mess}
\begin{split}
    A(k)\sim (k^2)^{m+1} + k^{2\alpha_1-1} \left[(k^2)^{m} + k^{2\alpha_2-1} \left[(k^2)^{m-1} + \cdots\right]\right],
\end{split}
\end{equation}
where we have dropped all prefactors independent of $k$ for simplicity. We can rewrite this in more compact notation as:
\begin{equation}
    A(k) \sim \sum_{r=0}^{m+1} A_i k^{2 r + \sum_{i \leq m+1-r} (2\alpha_i -1) }, 
\end{equation}
reproducing exactly the $(m+2)$ possible exponents of the reciprocal rule in Eq.~\eqref{eq:beta_mpole}. The dominant term for small $k$ is the one with the smallest exponent, which determines the dynamical exponent for relaxation $\beta^{(m)}$. If the dominant term is $k^\eta$ with $\eta \le 0$, then the hydrodynamic approximation breaks down, and the equation for $\rho(k,t)$ gives exponential (or faster) decay of the charge.

Now, if $\alpha_{i\leq n} < 1/2$ for some $n$, then all $n$ of the leftmost (dominant) integrals are IR divergent, and we must replace the infinite upper integration limit with the finite system size $L$. Then, these integrals yield an overall factor of $L^\nu$ (where $\nu$ is some positive power) that is canceled by the implicit Kac factor. The remaining $(m+1)-n$ integrals to the right are convergent even if we take the upper integration limit to infinity; $\int_1^{x_{n}} dx_{n+1} \cdots \rightarrow \int_1^{L} dx_{n+1} \cdots \rightarrow \int_1^{\infty} dx_{n+1} \cdots $. 

The problem, therefore, reduces to that of finding the scaling for a $[(m+1)-n]$-pole conserving system, which is exactly like setting all the previous $\tilde{\beta}_{i\leq n}^{-1} = 0$ in the reciprocal rule (see Table~\ref{tab:recip}) and ignoring their contribution to the overall dynamical exponent. This can also be motivated by a physical point of view because when the coupling between m-poles, $(m-1)$-poles, etc., becomes ultra-long-range, then it is as if these moments are no longer locally conserved. The local hydrodynamics thus effectively ``forgets'' about the higher multipole moments and reduces to the $[(m+1)-n]$-pole conserving case. For more complicated regimes where only a non-sequential subset of the $\{\alpha_i\}$ are less than 1/2, the dynamical exponents can be found by analyzing the divergences of \eqref{eq:general_prefactor} directly, and the reciprocal rule no longer exhibits an obvious hierarchy of moments.

\end{document}